\documentclass[12pt]{iopart}
\pdfoutput=1

\usepackage{nicefrac}
\usepackage{makecell} 		
\usepackage{rotating}					
\usepackage{amsfonts}

\usepackage{accents}		
\newcommand*{\dt}[1]{%
  \accentset{\mbox{\large\bfseries .}}{#1}}
  
\begin{document}

\title[Impulse Response Function for Brownian Motion]{Impulse Response Function for Brownian Motion}

\author{Nicos Makris}

\address{Dept. of Civil and Environmental Engineering, Southern Methodist University, Dallas, Texas, 75276}
\ead{nmakris@smu.edu}
\vspace{10pt}

\begin{abstract}
Motivated from the central role of the mean-square displacement and its second time-derivative --- that is the velocity autocorrelation function $\left\langle v(0)v(t)\right\rangle=\frac{1}{2} \frac{\mathrm{d}^{2} \left\langle  \Delta r^{2} (t)\right\rangle}{\mathrm{d}t^{2}} $ in the description of Brownian motion and its implications to microrheology, we revisit the physical meaning of the first time-derivative of the mean-square displacement of Brownian particles. By employing a rheological analogue for Brownian motion, we show that the time-derivative of the mean-square displacement $\frac{\mathrm{d}\left\langle \Delta r^{2} (t) \right\rangle}{\mathrm{d}t}$ of Brownian microspheres with mass $m$ and radius $R$ immersed in any linear, isotropic viscoelastic material is identical to $\frac{N K_B T}{3 \pi R}h(t)$, where $h(t)$ is the impulse response function $($strain history $\gamma (t)$, due to an impulse stress $\tau (t)=\delta (t-0))$ of a rheological network that is a parallel connection of the linear viscoelastic material with an inerter with distributed inertance $m_R=\frac{m}{6 \pi R}$. The impulse response function $h(t)=\frac{3\pi R}{N K_B T}\frac{\mathrm{d}\left\langle \Delta r^{2} (t) \right\rangle}{\mathrm{d}t}$ of the viscoelastic material--inerter parallel connection derived in this paper at the stress--strain level of the rheological analogue is essentially the response function $\chi(t)=\frac{h(t)}{6\pi R}$ of the Brownian particles expressed at the force--displacement level by Nishi \textit{et al.} \cite{NishiKilfoilSchmidtMacKintosh2018} after making use of the fluctuation--dissipation theorem. By employing the viscoelastic material--inerter rheological analogue we derive the mean-square displacement and its time-derivatives of Brownian particles immersed in a viscoelastic material described with a Maxwell element connected in parallel with a dashpot which captures the high-frequency viscous behavior and we show that for Brownian motion in such fluid-like soft matter the impulse response function, $h(t)$ maintains a finite constant value in the long term.
\end{abstract}

\section{Introduction}\label{sec:Sec01}

Soft materials such as colloidal dispersions, gels or polymer solutions exhibit a rich linear viscoelastic behavior and their dynamic response manifests several characteristic time-scales which are reflected in their bulk viscoleastic properties. For instance, the relaxation modulus $G_{\textit{ve}}(t)$ of a viscoelastic material is the resulting shear stress, $\tau(t)$ due to a unit-step shear strain, $\gamma(t)=U(t-0)$ \cite{Ferry1980,BirdArmstrongHassager1987}. All time response functions such as the relaxation modulus, $G_{\textit{ve}}(t)$ are causal functions --- that is they are zero at negative times, so their Fourier transform is essentially a Laplace transform. The Laplace transform of the relaxation modulus, $G_{\textit{ve}}(t)$ is the complex dynamic viscosity, $\eta_\textit{ve}(s)=\eta_1(s)+\mathrm{i}\eta_2(s)=\displaystyle \int_0^\infty G_{\textit{ve}}(t)e^{-st}\, \mathrm{d}t$ where $s=\mathrm{i}\omega$ is the Laplace variable and $\omega$ is the angular frequency. Traditional measurements of the complex dynamic viscosity $\eta_\textit{ve}(s)$ or the complex dynamic modulus, $\mathcal{G}_\textit{ve}(s)=s\eta_\textit{ve}(s)$ using rheometers are limited to the frequency range determined primarily by the inertia of the apparatus. With microrheology \cite{MasonWeitz1995,MasonGangWeitz1997,PalmerXuWirtz1998, SquiresMason2010,GittesSchnurrOlmstedMacKintoshSchmidt1997, SchnurrGittesMacKintoshSchmidt1997,GardelValentineWeitz2005,Waigh2005,LiKheifetsMedellinRaizen2010, HuangChavezTauteLukicJeneyRaizenFlorin2011,LiRaizen2013} the bulk viscoelastic characteristics of materials are inferred by monitoring the thermally-driven Brownian motion of probe microspheres suspended within the viscoelastic material and subjected to the perpetual random forces from the collisions of the molecules of the material. The thermal fluctuations of the immersed microparticles have been monitored with dynamic light scattering (DLS) and diffusing wave spectroscopy (DWS) \cite{MasonWeitz1995, MasonGangWeitz1997, PalmerXuWirtz1998, SquiresMason2010} or with laser interferometry \cite{GittesSchnurrOlmstedMacKintoshSchmidt1997,SchnurrGittesMacKintoshSchmidt1997,GardelValentineWeitz2005,Waigh2005, LiKheifetsMedellinRaizen2010,HuangChavezTauteLukicJeneyRaizenFlorin2011,LiRaizen2013} with nanometer spatial resolution and submicrosecond temporal resolution; allowing measuring frequency response at frequencies that far exceed the limitations of mechanical rheometers.

The dynamics of Brownian motion have been traditionally expressed with the mean-square displacement,
\begin{equation}\label{eq:Eq01}
\left\langle \Delta r^{2} (t) \right\rangle = \frac{1}{M} \sum_{j=1}^{M}\left(r_j(t)-r_j(0)\right)^{2}
\end{equation}
where $M$ is the number of suspended microparticles; while $r_j(t)$ and $r_j(0)$ are the positions of particle $j$ at time $t$ and the time origin, $t=$ 0; in association with the velocity autocorrelation function of the Brownian particles
\begin{equation}\label{eq:Eq02}
\left\langle v(0) v(t) \right\rangle  = \left\langle v(\xi) v(\xi+t) \right\rangle = \lim_{T\rightarrow\infty} \frac{1}{T} \int_{0}^T v(\xi) v(\xi+t) \, \mathrm{d}\xi
\end{equation}
where $v(t)$ is the velocity of the Brownian particle \cite{Langevin1908,UhlenbeckOrnstein1930, WangUhlenbeck1945, Attard2012, KalmykovCoffey2017}. 

The Laplace transform of the mean-square displacement, $\mathcal{L}\left\lbrace \left\langle \Delta r^{2} (t) \right\rangle \right\rbrace = \left\langle \Delta r^{2} (s) \right\rangle = \displaystyle \int_{0}^\infty \left\langle\Delta r^{2} (t) \right\rangle e^{-st} \, \mathrm{d}t$ is related with the Laplace transform of the velocity autocorrelation function $\mathcal{L}\left\lbrace \left\langle v(0) v(t) \right\rangle \right\rbrace = \left\langle v(0) v(s)\right\rangle = \displaystyle \int_{0}^\infty \left\langle v(0) v(t) \right\rangle e^{-st} \, \mathrm{d}t$ via the identity \cite{SquiresMason2010,Attard2012, KalmykovCoffey2017}
\begin{equation}\label{eq:Eq03}
\left\langle v(0) v(s)\right\rangle = \frac{s^{2}}{2} \left\langle \Delta r^{2} (s) \right\rangle
\end{equation}
while, according to the properties of the Laplace transform of the derivatives of a function
\begin{equation}\label{eq:Eq04}
s^{2} \left\langle \Delta r^{2} (s) \right\rangle  = \mathcal{L} \left\lbrace \frac{\mathrm{d}^{2}\left\langle \Delta r^{2} (t) \right\rangle}{\mathrm{d}t^{2}} \right\rbrace + s\left\langle \Delta r^{2} (0) \right\rangle + \frac{\mathrm{d}\left\langle \Delta r^{2} (0) \right\rangle}{\mathrm{d}t}
\end{equation}
From the definition of the mean-square displacement given by Eq. (\ref{eq:Eq01}), at the time origin $t=$ 0, $\left\langle \Delta r^{2} (0) \right\rangle=$ 0. Furthermore, the time-derivative of Eq. (\ref{eq:Eq01}) gives
\begin{equation}\label{eq:Eq05}
\frac{\mathrm{d}\left\langle \Delta r^{2} (t) \right\rangle}{\mathrm{d}t}=\frac{2}{M}\sum_{j=1}^{M}\left(r_j(t)-r_j(0)\right) \frac{\mathrm{d}r_j(t)}{\mathrm{d}t}
\end{equation}
therefore, at $t=$ 0, $\frac{\mathrm{d}\left\langle \Delta r^{2} (t) \right\rangle}{\mathrm{d}t}=$ 0. Accordingly,  substitution of Eq. (\ref{eq:Eq04}) into Eq. (\ref{eq:Eq03}) gives
\begin{equation}\label{eq:Eq06}
\mathcal{L} \left\lbrace \left\langle v(0) v(t)\right\rangle \right\rbrace = \frac{1}{2} \mathcal{L} \left\lbrace \frac{\mathrm{d}^{2}\left\langle \Delta r^{2} (t) \right\rangle}{\mathrm{d}t^{2}} \right\rbrace 
\end{equation}
and inverse Laplace transform of Eq. (\ref{eq:Eq06}) yields
\begin{equation}\label{eq:Eq07}
\left\langle v(0) v(t)\right\rangle = \frac{1}{2} \frac{\mathrm{d}^{2}\left\langle \Delta r^{2} (t) \right\rangle}{\mathrm{d}t^{2}}
\end{equation}
which shows that the velocity autocorrelation function is half the second time-derivative of the mean-square displacement \cite{KenkreKuhneReineker1981,BianKimKarniadakis2016}.

The phenomenon of Brownian motion was first explained in the 1905 Einstein's celebrated paper \cite{Einstein1905} which examined the long-term response of Brownian microspheres with mass $m$ and radius $R$ suspended in a memoryless, Newtonian fluid with viscosity $\eta$. Einstein's theory of Brownian motion predicts the long-term expression for the mean-square displacement of the randomly moving microspheres (diffusive regime)
\begin{equation}\label{eq:Eq08}
\left\langle \Delta r^{2} (t) \right\rangle = 2 N D t = \frac{N K_B T}{3 \pi R} \frac{1}{\eta} t
\end{equation}
where $N\in \left\lbrace 1,2,3 \right\rbrace$ is the number of spatial dimensions, $K_B$ is Boltzman's constant, $T$ is the equilibrium temperature of the Newtonian fluid with viscosity $\eta$ within which the Brownian microspheres are immersed and $D=\frac{K_B T}{6 \pi R \eta}$ is the diffusion coefficient. The time derivative of Eq. (\ref{eq:Eq08}), $\frac{\mathrm{d}\left\langle \Delta r^{2}(t) \right\rangle}{\mathrm{d}t}=2 N D$ is a constant which is in contradiction with the result of Eq. (\ref{eq:Eq05}) at the time origin $(t=0)$.

At short time-scales \cite{LiKheifetsMedellinRaizen2010,HuangChavezTauteLukicJeneyRaizenFlorin2011,LiRaizen2013}, when $t<\frac{m}{6\pi R \eta}= \tau$, the Brownian motion of suspended particles is influenced by the inertia of the particle and the surrounding fluid (ballistic regime); and Einstein's ``long-term'' result offered by Eq. (\ref{eq:Eq08}) was extended for all time-scales by \cite{UhlenbeckOrnstein1930}
\begin{equation}\label{eq:Eq09}
\left\langle \Delta r^{2} (t) \right\rangle = \frac{N K_B T}{3 \pi R} \frac{1}{\eta} \left[ t-\tau \left(1-e^{-\nicefrac{t}{\tau}} \right) \right]
\end{equation}
where $\tau=\frac{m}{6\pi R \eta}$ is the dissipation time-scale of the perpetual fluctuation--dissipation process. The time derivative of Eq. (\ref{eq:Eq09}) is
\begin{equation}\label{eq:Eq10}
\frac{\mathrm{d}\left\langle \Delta r^{2}(t) \right\rangle}{\mathrm{d}t} = \frac{N K_B T}{3 \pi R} \frac{1}{\eta} \left(1-e^{-\nicefrac{t}{\tau}}\right)
\end{equation}
indicating that at $t=$ 0, $\frac{\mathrm{d}\left\langle \Delta r^{2}(t) \right\rangle}{\mathrm{d}t} =$ 0 which is in agreement with the result of Eq. (\ref{eq:Eq05}). Equation (\ref{eq:Eq07}) in association with the result of Eq. (\ref{eq:Eq09}) yields the velocity autocorrelation function of Brownian particles with mass $m$ when suspended in a memoryless, Newtonian fluid with viscosity $\eta$
\begin{equation}\label{eq:Eq11}
\left\langle v(0)v(t) \right\rangle = \frac{1}{2} \frac{\mathrm{d}^{2}\left\langle \Delta r^{2}(t) \right\rangle}{\mathrm{d}t^{2}} = \frac{N K_B T}{m} e^{-\nicefrac{t}{\tau}}
\end{equation}
which is the classical result derived by \cite{UhlenbeckOrnstein1930} after evaluating ensemble averages of the random Brownian process. Equation (\ref{eq:Eq11}); while valid for all time-scales it does not account for the hydrodynamic memory that manifests as the energized Brownian particle displaces the fluid in its immediate vicinity \cite{ZwanzigBixon1970, Widom1971, Hinch1975,  ClercxSchram1992, Franosch_etal2011,JannaschMahamdehSchaffer2011}.

The reader recognizes that the exponential term of the velocity autocorrelation function given by Eq. (\ref{eq:Eq11}) is whatever is left after taken the second time derivative of the mean-square displacement given by Eq. (\ref{eq:Eq09}) that is valid for all time-scales. Consequently, by accounting for the ``ballistic regime'' at short time-scales, Uhlenbeck and Ornstein's (1930) expression for the mean-square displacement given by Eq. (\ref{eq:Eq09}) is consistent with the identity given by Eq. (\ref{eq:Eq07}) and yields the correct expression for the velocity autocorrelation function of Brownian particles suspended in a memoryless Newtonian fluid given by Eq. (\ref{eq:Eq11}). In contrast, Einstein's (1905) ``long-term'' expression for the mean-square displacement given by Eq. (\ref{eq:Eq08}) (diffusive regime) yields an invariably zero velocity autocorrelation function.

Studies on the behavior of hard-sphere systems have identified a $\frac{1}{t}$ decay of $\frac{\mathrm{d}\left\langle \Delta r^{2}(t) \right\rangle}{\mathrm{d}t}$ with time \cite{Sperl2005}; while with reference to Eq. (\ref{eq:Eq08}), the time-derivative of the mean-square displacement, $\frac{\mathrm{d}\left\langle \Delta r^{2}(t) \right\rangle}{\mathrm{d}t}$, has been interpreted as a time-dependent diffusion coefficient \cite{SegrePusey1996}. Given that the mean-square displacement defined by Eq. (\ref{eq:Eq01}) and its second time derivative --- that is the velocity autocorrelation function defined by Eq. (\ref{eq:Eq02}), play such a central role in the description of Brownian motion in association with the role of $\frac{\mathrm{d}\left\langle \Delta r^{2}(t) \right\rangle}{\mathrm{d}t}$ to interpret the Brownian motion at various time scales \cite{KhanMason2014,KhanMason2014PRE} and confined spacings \cite{GhoshKrishnamurthy2018}; in this paper we revisit the physical meaning of the first time derivative of the mean-square displacement, $\frac{\mathrm{d}\left\langle \Delta r^{2}(t) \right\rangle}{\mathrm{d}t}$ by employing the viscous--viscoelastic correspondence principle for Brownian motion \cite{Makris2020}.
\begin{figure*}[t!]
\centering
\includegraphics[width=.75\linewidth, angle=0]{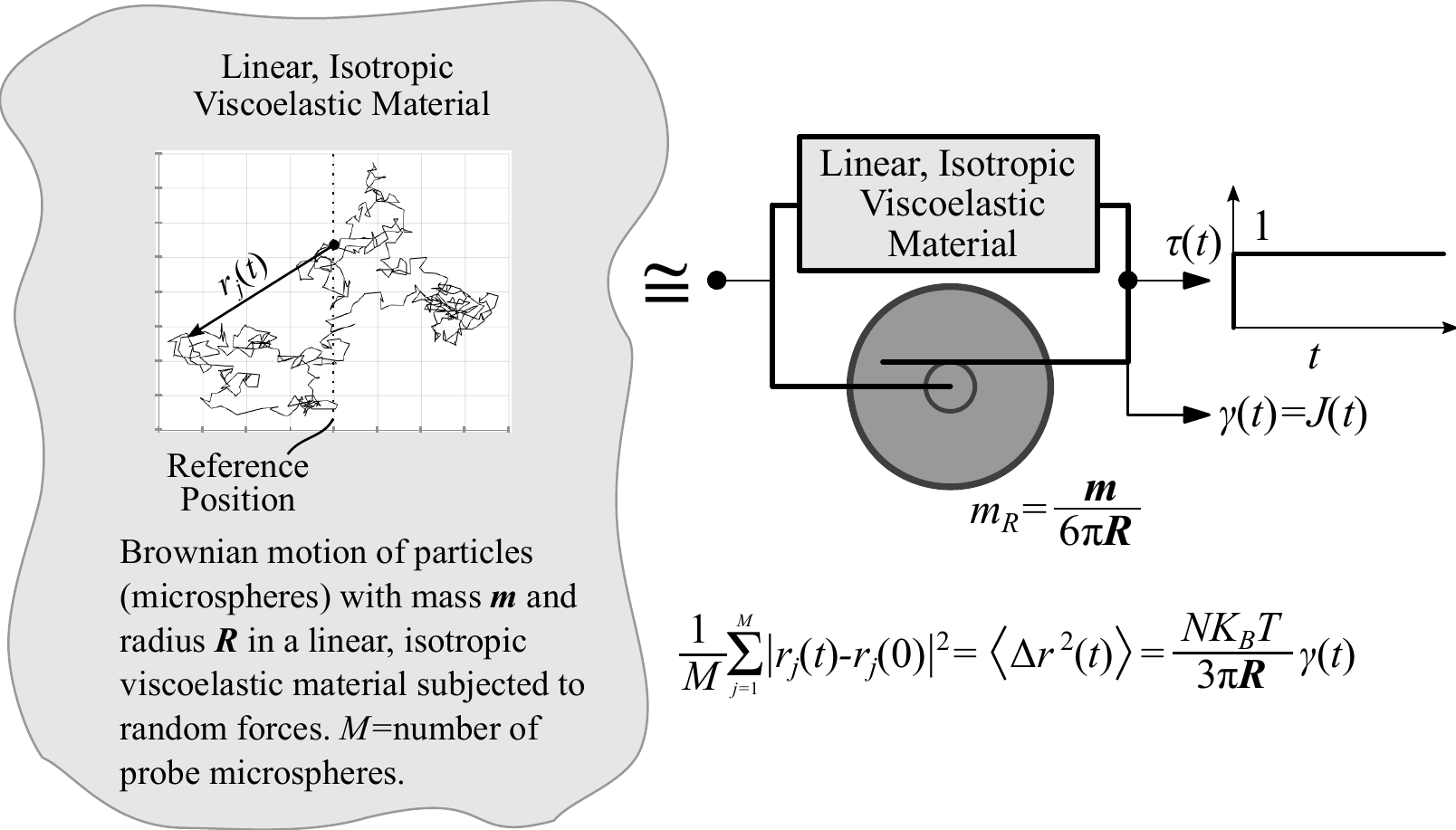}
\caption{Statement of the viscous--viscoelastic correspondence principle for Brownian motion \cite{Makris2020}. The mean-square displacement, $\left\langle \Delta r^{2}(t) \right\rangle$, of Brownian particles (microspheres) with mass $m$ and radius $R$ suspended in any linear, isotropic viscoelastic material when subjected to the random forces from the collisions of the molecules of the viscoelastic material, is identical to {\large $\frac{NK_B T}{3\pi R}$}$\gamma (t)$, where $\gamma (t)=J(t)$ is the strain due to a unit step--stress on a rheological network that is a parallel connection of the linear viscoelastic material and an inerter with distributed inertance $m_R=$ {\large $\frac{m}{6\pi R}$}.}
\label{fig:Fig01}
\end{figure*}

\section{A Rheological Analogue for Brownian Motion}\label{sec:Sec02}

In a recent publication, Makris \cite{Makris2020} presented a viscous--viscoelastic correspondence principle for Brownian motion which reveals that the mean-square displacement, $\left\langle \Delta r^{2}(t) \right\rangle$ of Brownian microspheres with mass $m$ and radius $R$ when suspended in any linear isotropic, viscoelastic material and subjected to the random forces from the collisions of the molecules of the viscoelastic material is identical to $\frac{N K_B T}{3 \pi R} \gamma (t)$, where $\gamma (t) = J (t)$ is the strain due to a unit-step stress on a rheological network that is a parallel connection of the linear viscoelastic material (within which the microspheres are immersed) with an inerter with distributed inertance $m_R = \frac{m}{6 \pi R}$. Accordingly,
\begin{equation}\label{eq:Eq12}
\left\langle \Delta r^{2} (t) \right\rangle = \frac{N K_B T}{3 \pi R} J (t)
\end{equation}
where $J(t)$ is the creep compliance of the rheological network shown in Fig. \ref{fig:Fig01} (right). Laplace transform of Eq. (\ref{eq:Eq12}) gives \cite{Makris2020}
\begin{equation}\label{eq:Eq13}
\left\langle \Delta r^{2} (s) \right\rangle = \frac{N K_B T}{3 \pi R}\mathcal{C}(s) = \frac{N K_B T}{3 \pi R} \frac{\mathcal{J} (s)}{s}
\end{equation}
where $\mathcal{C}(s)$ is the complex creep function \cite{PalmerXuWirtz1998, EvansTassieriAuhlWaigh2009, Makris2019} and $\mathcal{J} (s) = \frac{1}{\mathcal{G} (s)}$ is the complex dynamic compliance of the rheological network shown in Fig. \ref{fig:Fig01} (right). The complex dynamic compliance, $\mathcal{J}  (s)$ of a rheological network is the inverse of the complex dynamic modulus, $\mathcal{G}  (s)$; and is a transfer function that relates a strain output, $\gamma (s)$ to a stress input $\tau (s)$ \cite{Ferry1980, BirdArmstrongHassager1987, Tschoegl1989}. In structural mechanics, the equivalent of the complex dynamic compliance at the displacement--force level is known as the dynamic flexibility, often expressed with $\mathcal{H} (\omega) = \frac{1}{\mathcal{K} (\omega)}$ \cite{CloughPenzien1970}, where $\mathcal{K} (\omega)$ is the dynamic stiffness of the structure. The Laplace transform of the time-derivative of the mean-square displacement is:
\begin{equation}\label{eq:Eq14}
\mathcal{L} \left\lbrace \frac{\mathrm{d}\left\langle \Delta r^{2} (t) \right\rangle}{\mathrm{d}t} \right\rbrace = s \left\langle \Delta r^{2} (s) \right\rangle - \left\langle \Delta r^{2} (0) \right\rangle
\end{equation}
From Eq. (\ref{eq:Eq01}), at the time origin, $t =$ 0, $\left\langle \Delta r^{2} (0) \right\rangle =$ 0, and substitution of Eq. (\ref{eq:Eq13}) into Eq. (\ref{eq:Eq14}) gives:
\begin{equation}\label{eq:Eq15}
\mathcal{L} \left\lbrace \frac{\mathrm{d}\left\langle \Delta r^{2} (t) \right\rangle}{\mathrm{d}t} \right\rbrace = \frac{N K_B T}{3 \pi R} \mathcal{J}  (s)
\end{equation}
The inverse Laplace transform of the complex dynamic compliance, $\mathcal{J} (s)$, appearing in the right-hand side of Eq. (\ref{eq:Eq15}) is the impulse fluidity, $\phi (t) = \mathcal{L}^{-1} \left\lbrace \mathcal{J} (s) \right\rbrace$ \cite{Giesekus1995, MakrisKampas2009, MakrisEfthymiou2020}, defined as the resulting strain $\gamma (t)$, due to an impulsive stress input, $\tau (t) = \delta (t - 0)$. The impulse function $\delta(t-0)$, is the Dirac delta function \cite{Lighthill1958} with the property $\mathcal{L} \left\lbrace \delta(t-\xi) \right\rbrace = \displaystyle\int^{t}_{0^-} \delta(t-\xi)e^{-st}\,\mathrm{d}t=e^{-\xi s} $. The equivalent of the impulse fluidity, $\phi (t)$, at the displacement--force level is the impulse response function often expressed as $h (t)$. Given that the term ``impulse response function'' (rather than the term ``impulse fluidity'') is widely known and used in dynamics \cite{CloughPenzien1970}, structural mechanics \cite{HarrisCrede1976}, electrical signal processing \cite{OppenheimSchafer1975,Reid1983} and economics \cite{BorovivckaHansenScheinkman2014,BorovivckaHansen2016}; in this paper we adopt the term ``\textit{impulse response function }$= h (t)$'', rather than the term ``impulse fluidity'' used narrowly in the viscoelasticity literature alone. Accordingly, inverse Laplace transform of Eq. (\ref{eq:Eq15}) gives:
\begin{equation}\label{eq:Eq16}
\frac{\mathrm{d}\left\langle \Delta r^{2} (t) \right\rangle}{\mathrm{d}t} = \frac{N K_B T}{3 \pi R} \mathcal{L}^{-1} \left\lbrace \mathcal{J} (s) \right\rbrace = \frac{N K_B T}{3 \pi R} h (t)
\end{equation}
Equation (\ref{eq:Eq16}) indicates that the time-derivative of the mean-square displacement, $\frac{\mathrm{d}\left\langle \Delta r^{2} (t) \right\rangle}{\mathrm{d}t}$, of Brownian particles suspended in any linear, isotropic viscoelastic material is proportional to the impulse response function, $h (t)$, of the rheological network shown in Fig. \ref{fig:Fig01} (right), defined as the resulting strain history, $\gamma (t)$, of the viscoelastic material--inerter parallel connection due to an impulsive stress input, $\tau (t) = \delta (t - 0)$. The result presented by Eq. (\ref{eq:Eq16}) has been reached by Nishi \textit{et al.} \cite{NishiKilfoilSchmidtMacKintosh2018} by first employing the fluctuation--dissipation theorem and relating the response function $\chi(t)$ of Brownian particles at the force--displacement level with the time-derivative of the position autocorrelation function $\left\langle r(0)r(t) \right\rangle$, and subsequently using the identity that relates the mean-square displacement $\left\langle \Delta r^2 (t) \right\rangle$ to the position autocorrelation function $\left\langle r(0)r(t) \right\rangle$. Accordingly, the relation of the response function $\chi(t)$ of Brownian particles at the force--displacement level introduced by Nishi \textit{et al.}\cite{NishiKilfoilSchmidtMacKintosh2018} and the impulse response $h(t)$ of the rheological analogue for the Brownian motion shown on the right of Fig. \ref{fig:Fig01} is $\chi(t)=\frac{h(t)}{6\pi R}$.

As an example, the corresponding rheological network for Brownian motion of microspheres with mass $m$ and radius $R$ immersed in a memoryless Newtonian fluid with viscosity $\eta$ is a dashpot--inerter parallel connection with constitutive law \cite{Makris2020}
\begin{equation}\label{eq:Eq17}
\tau (t) = \eta \frac{\mathrm{d} \gamma (t)}{\mathrm{d}t} + m_R \frac{\mathrm{d}^2 \gamma (t)}{\mathrm{d}t^2}
\end{equation}
where $m_R = \frac{m}{6 \pi R}$ is the distributed inertance of the inerter with units [M][L]$^{-1}$ (i.e. Pa s$^2$). The Laplace transform of Eq. (\ref{eq:Eq17}) gives:
\begin{equation}\label{eq:Eq18}
\tau (s) = \mathcal{G} (s) \gamma (s) = (\eta s + m_R s^2) \gamma (s)
\end{equation}
where $\mathcal{G} (s) = \frac{1}{\mathcal{J} (s)} = \eta s + m_R s^2$ is the complex dynamic modulus of the dashpot--inerter parallel connection (inertoviscous fluid, \cite{Makris2017}). The complex dynamic compliance, $\mathcal{J} (s)$, of the inertoviscous fluid expressed by Eq. (\ref{eq:Eq17}) is:
\begin{equation}\label{eq:Eq19}
\mathcal{J} (s) = \frac{1}{\mathcal{G} (s)} = \frac{1}{\eta s + m_R s^2} = \frac{1}{\eta} \left( \frac{1}{s} - \frac{1}{s + \frac{1}{\tau}} \right)
\end{equation}
where $\tau = \frac{m_R}{\eta} = \frac{m}{6 \pi R \eta}$ is the dissipation time , which is the time scale needed for the kinetic energy stored in the inerter with distributed inertance, $m_R$, to be dissipated by the dashpot with viscosity, $\eta$. Inverse Laplace transform of Eq. (\ref{eq:Eq19}) gives:
\begin{equation}\label{eq:Eq20}
\mathcal{L}^{-1} \left\lbrace \mathcal{J} (s) \right\rbrace = h (t) = \frac{1}{\eta} ( 1 - e^{-\nicefrac{t}{\tau}} )
\end{equation}
By substitution of the result of Eq. (\ref{eq:Eq20}) into Eq. (\ref{eq:Eq16}) we recover Eq. (\ref{eq:Eq10}) that was reached by merely taking the time-derivative of Eq. (\ref{eq:Eq09}), initially derived by Uhlenbeck and Ornstein \cite{UhlenbeckOrnstein1930} after computing ensemble averages of the random Brownian process.

The analysis presented in this section, in association with the viscous--viscoelastic correspondence principle for Brownian motion \cite{Makris2020} concludes that the time-derivative of the mean-square displacement, $\frac{\mathrm{d}\left\langle \Delta r^{2} (t) \right\rangle}{\mathrm{d}t}$, of Brownian microspheres with mass $m$ and radius $R$ suspended in any linear viscoelastic material is identical to $\frac{N K_B T}{3 \pi R} h (t)$, where $h (t)$ is the impulse response function of a rheological network that is a parallel connection of the linear viscoelastic material and an inerter with distributed inertance $m_R = \frac{m}{6 \pi R}$.

\section{Impulse Response Function for Brownian Motion in a Harmonic Trap (Kelvin--Voigt Solid)}\label{sec:Sec03}

The Brownian motion of microparticles trapped in a harmonic potential when excited by random forces $f_R (t)$ has been studied by \cite{UhlenbeckOrnstein1930} and \cite{WangUhlenbeck1945}. The mean-square displacement of a Brownian particle in a harmonic trap has been evaluated by \cite{WangUhlenbeck1945} after computing the velocity autocorrelation function of the random process. For the underdamped case ($\omega_0 \tau > \frac{1}{2}$):
\begin{equation}\label{eq:Eq21}
\left\langle \Delta r^{2} (t) \right\rangle  = \frac{2 N K_B T}{m \omega_0 ^2} \left[ 1 - e^{-\nicefrac{t}{2 \tau}} \left( \cos (\omega_D t) + \frac{1}{2 \omega_D \tau} \sin (\omega_D t) \right) \right]
\end{equation}
where $\omega_0 = \sqrt{\frac{k}{m}}$ is the undamped natural frequency of the trapped particle with mass $m$ and radius $R$, $\tau = \frac{m}{6 \pi R \eta}$ is the dissipation time and $\omega_D = \omega_0 \sqrt{1 - \left( \frac{1}{2 \omega_0 \tau} \right)}$ is the damped angular frequency of the trapped particle \cite{LiKheifetsMedellinRaizen2010,LiRaizen2013}.

\begin{figure}[t!]
\centering
\includegraphics[width=.6\linewidth, angle=0]{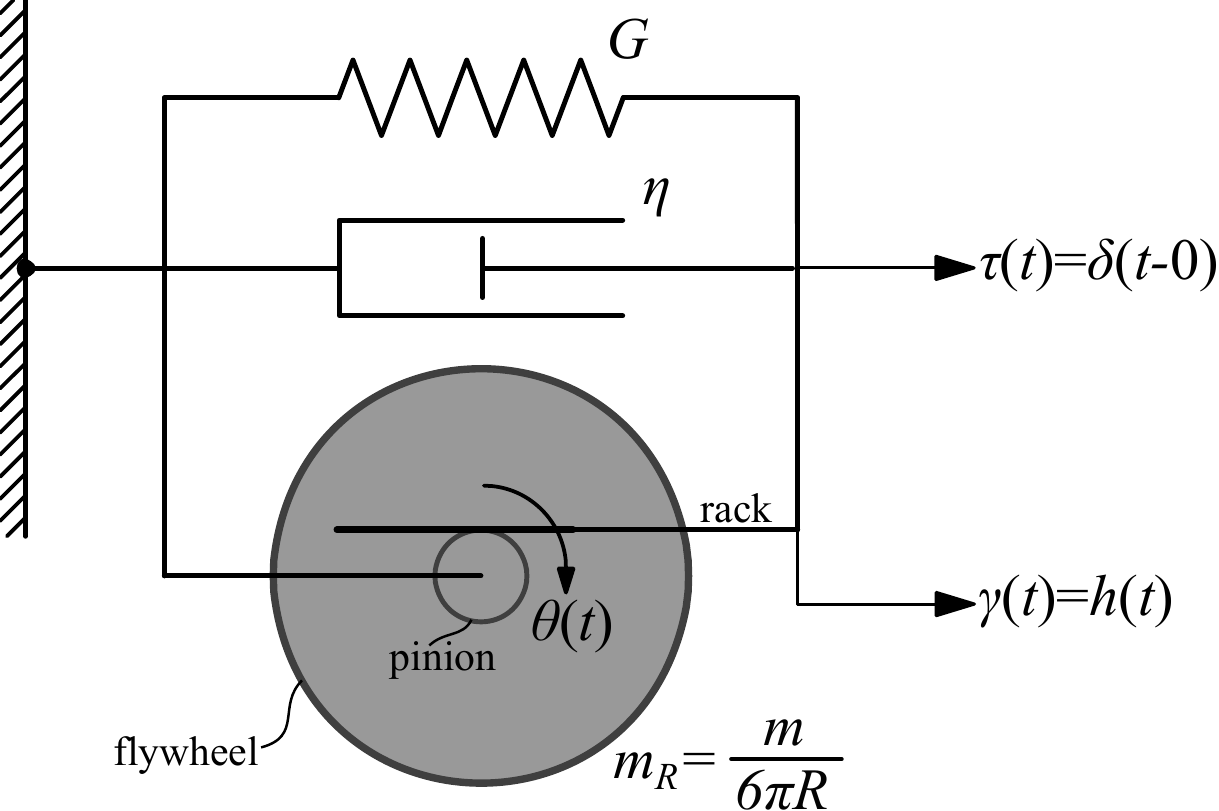}
\caption{Inertoviscoelastic solid which is a parallel connection of an inerter with a distributed inertance $m_R$, a dashpot with viscosity $\eta$ and a linear spring with elastic shear modulus $G$. In analogy with the traditional schematic of a dashpot that is a hydraulic piston, the distributed inerter is depicted schematically with a rack--pinion--flywheel system.}
\label{fig:Fig02}
\end{figure}

The rheological analogue for Brownian motion of particles trapped in a Kelvin--Voigt solid is the inertoviscoelastic solid shown in Fig. \ref{fig:Fig02} which is a parallel connection of a spring with elastic shear modulus $G$ and a dashpot with shear viscosity $\eta$ (Kelvin--Voigt soild); together with an inerter with distributed inertance $m_R = \frac{m}{6 \pi R}$. Given the parallel connection of the three elementary mechanical elements shown in Fig. \ref{fig:Fig02}, the constitutive law of the combined inertoviscoelastic solid is \cite{Makris2020}:
\begin{equation}\label{eq:Eq22}
\tau (t) = G \gamma (t) + \eta \frac{\mathrm{d} \gamma (t)}{\mathrm{d}t} + m_R \frac{\mathrm{d}^2 \gamma (t)}{\mathrm{d}t^2}
\end{equation}
The Laplace transform of Eq. (\ref{eq:Eq22}) is:
\begin{equation}\label{eq:Eq23}
\tau (s) = \mathcal{G} (s) \gamma (s) = (G + \eta s + m_R s^2) \gamma (s)
\end{equation}
where $\mathcal{G} (s) = G + \eta s + m_R s^2$ is the complex dynamic modulus; while the complex dynamic compliance of the inertoviscoelastic solid is:
\begin{equation}\label{eq:Eq24}
\mathcal{J} (s) = \frac{1}{\mathcal{G} (s)}  = \frac{1}{G + \eta s + m_R s^2}  = \frac{1}{m_R} \frac{1}{\left( s + \frac{1}{2 \tau} \right)^2 + \omega_R^2 - \left( \frac{1}{2 \tau} \right)^2}
\end{equation}
where again $\tau = \frac{m_R}{\eta} = \frac{m}{6 \pi R \eta }$ is the dissipation time and $\omega_R = \sqrt{\frac{G}{m_R}}$ is the undamped rotational angular frequency of the inertoviscoelastic solid shown in Fig. \ref{fig:Fig02}. For the inertoviscoelastic solid described by Eq. (\ref{eq:Eq22}) and a Brownian particle in a harmonic trap to have the same undamped natural frequency $\omega_R = \sqrt{\frac{G}{m_R}} = \sqrt{\frac{k}{m}} = \omega_0$, the shear modulus needs to assume the value $G = \frac{k}{6 \pi R}$, where $k$ is the spring constant of the harmonic trap \cite{WangUhlenbeck1945, LiRaizen2013}. Accordingly, by setting $\omega_R = \omega_0$, the last two terms in the denominator of Eq. (\ref{eq:Eq24}) combine to $\omega_R^2 \left[ 1 - \left( \frac{1}{2 \omega_R \tau} \right)^2 \right] = \omega_0^2 \left[ 1 - \left( \frac{1}{2 \omega_0 \tau} \right)^2 \right] = \omega_D^2$.

\begin{figure}[t!]
\centering
\includegraphics[width=.7\linewidth, angle=0]{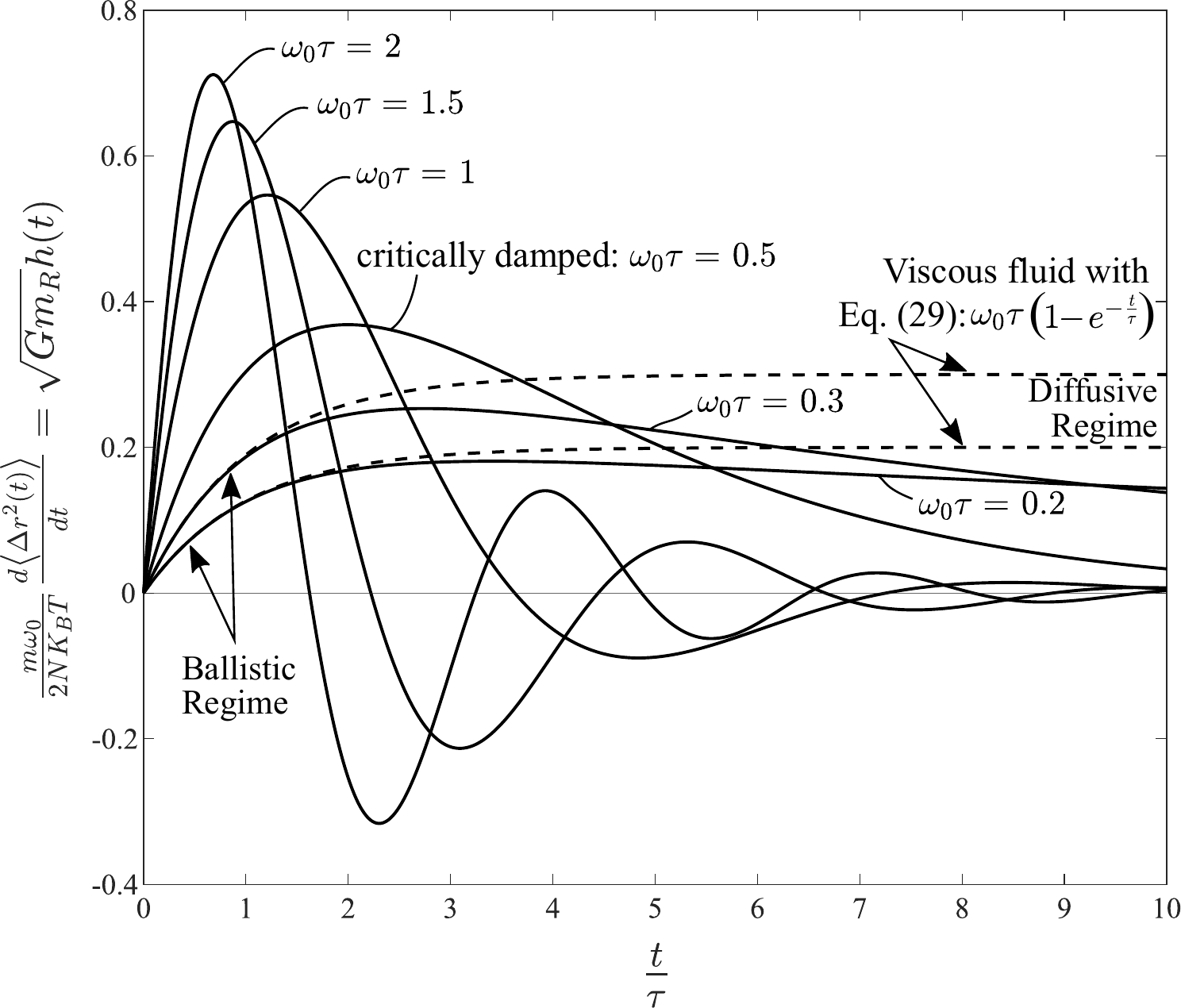}
\caption{Normalized time-derivative of the mean-square displacement of Brownian particles trapped in a harmonic potential for the underdamped, $\omega_0 \tau>\frac{1}{2}$, critically damped, $\omega_0 \tau=\frac{1}{2}$, and overdamped cases which is equal to $\sqrt{G m_R}h(t)$ of the inertoviscoelastic solid shown in Fig. \ref{fig:Fig02}. For the overdamped cases (weak spring) at early times (ballistic regime), the time--response functions of the Brownian particles in a harmonic trap coincide with the corresponding time--response functions of Brownian particles in a viscous fluid with viscosity $\eta$.}
\label{fig:Fig03}
\end{figure}

The inverse Laplace transform of Eq. (\ref{eq:Eq24}) yields the impulse response function of the inertoviscoelastic solid described by Eq. (\ref{eq:Eq22}) \cite{Makris2018}
\begin{equation}\label{eq:Eq25}
\mathcal{L}^{-1} \left\lbrace \mathcal{J} (s) \right\rbrace = h (t) = \frac{1}{m_R} \frac{1}{\omega_D} e^{-\nicefrac{t}{2 \tau}} \sin (\omega_D t)
\end{equation}
Substitution of the result of Eq. (\ref{eq:Eq25}) into Eq. (\ref{eq:Eq16}), yields that the time derivative of the mean-square displacement of Brownian particles trapped in a Kelvin--Voigt solid is:
\begin{equation}\label{eq:Eq26}
\frac{\mathrm{d}\left\langle \Delta r^{2} (t) \right\rangle}{\mathrm{d}t}  = \frac{N K_B T}{3 \pi R} h (t)  = \frac{2 N K_B T}{m \omega_0} \frac{1}{\sqrt{1 - \left( \frac{1}{2 \omega_0 \tau} \right)^2}} e^{-\nicefrac{t}{2 \tau}} \sin (\omega_D t)
\end{equation}
The result of Eq. (\ref{eq:Eq26}) that was computed herein after calculating the impulse response function of the inertoviscoelastic solid in association with Eq. (\ref{eq:Eq16}) is identical to the first time derivative of Eq. (\ref{eq:Eq21}) derived by \cite{WangUhlenbeck1945} after computing ensemble averages of the random Brownian process.

In a dimensionless form Eq. (\ref{eq:Eq26}) or (\ref{eq:Eq25}) which is for the underdamped case $\left( \omega_0 \tau > \frac{1}{2} \right)$ is expressed as:
\begin{eqnarray}\label{eq:Eq27}
\frac{m \omega_0}{2 N K_B T}  \frac{\mathrm{d}\left\langle \Delta r^{2} (t) \right\rangle}{\mathrm{d}t} & = \sqrt{G m_R} h (t) \\ \nonumber 
& = \frac{1}{\sqrt{1 - \left( \frac{1}{2 \omega_0 \tau} \right)^2}} e^{-\nicefrac{t}{2 \tau}} \sin \left(\omega_0 \tau \sqrt{1 - \left( \frac{1}{2 \omega_0 \tau} \right)^2} \frac{t}{\tau} \right)
\end{eqnarray}
For the overdamped case $\left( \omega_0 \tau < \frac{1}{2} \right)$, the normalized impulse response function for Brownian motion of particles trapped in a Kelvin--Voigt solid is:
\begin{sidewaystable}
\caption{Mean-square displacement together with its first and second time derivative (autocorrelation function) of Brownian microspheres with mass $m$ and radius $R$ suspended in a viscous Newtonian fluid, a Kelvin--Voigt solid, a Maxwell fluid and a subdiffusive Scott--Blair fluid. The mechanical analogues for Brownian motion in these materials are shown in the first column, whereas the deterministic expressions of their creep compliances, $J(t)$, impulse response functions, $h(t)$ and strain-rate response functions, $\psi(t)$ are shown in the subsequent columns.}
\scriptsize
\begin{tabular*}{\textwidth}{llll}
\br
	\thead[l]{Brownian Motion \\ of microspheres with \\ mass $m$ and radius \\ $R$ suspended in a:} & \thead[l]{$\left\langle \Delta r^{2} (t) \right\rangle = \frac{N K_B T}{3 \pi R} J(t) $ \\ \\ $J(t)=\mathcal{L}^{-1}\left\lbrace \mathcal{C}(s) \right\rbrace=$ {\scriptsize Creep Compliance} \\ $\mathcal{C}(s)=\mathcal{L}\left\lbrace J(t) \right\rbrace=$ {\scriptsize Complex Creep Function}} & \thead[l]{$\frac{\mathrm{d}\left\langle \Delta r^{2} (t) \right\rangle}{\mathrm{d}t} = \frac{N K_B T}{3 \pi R} h(t) $ \\ \\  $h(t)=\mathcal{L}^{-1}\left\lbrace \mathcal{J}(s) \right\rbrace=$ Impulse Response \\ \quad \quad \quad \quad \quad \quad \quad \quad \quad \quad \quad \quad \quad \quad Function \\ $\mathcal{J}(s)=\mathcal{L}\left\lbrace h(t) \right\rbrace=$ Complex Dynamic \\ \quad \quad \enskip Compliance (Dynamic Flexibility)} & \thead[l]{\quad $\frac{1}{2} \frac{\mathrm{d}^{2}\left\langle \Delta r^{2}(t) \right\rangle}{\mathrm{d}t^{2}}=\left\langle v(0)v(t) \right\rangle $ \\ $= \displaystyle \lim_{T\rightarrow\infty} \frac{1}{T} \displaystyle\int_{0}^T v(\xi) v(\xi+t) \, \mathrm{d}\xi = \frac{N K_B T}{6 \pi R} \psi(t)$ \\ $\psi(t)=\mathcal{L}^{-1}\left\lbrace \mathcal{\phi}(s) \right\rbrace=$ Impulse Strain-Rate \\ \quad \quad \quad \quad \quad \quad \quad \quad \quad \quad  Response Function \\ $\mathcal{\phi}(s)=\mathcal{L}\left\lbrace \psi(t) \right\rbrace =$ Complex Dynamic Fluidity \\ \quad \quad \quad \quad \quad \quad \quad \quad \quad \quad \quad \quad  (Admittance)} \\
\mr
\thead[l]{Newtonian Viscous Fluid\\with viscosity $\eta$\\ \includegraphics[scale=0.30]{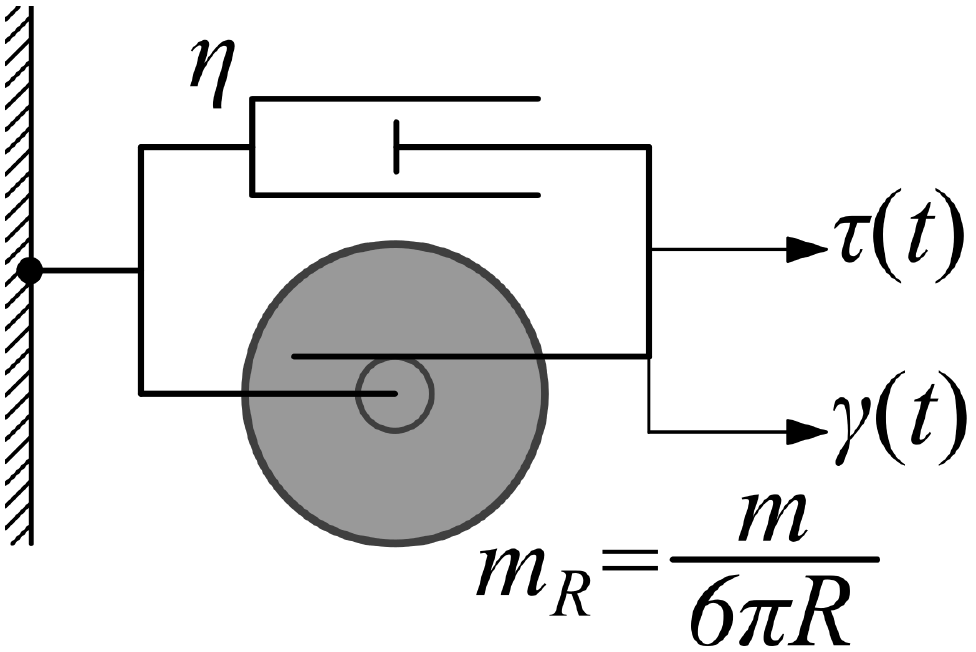}} & {\scriptsize $J(t)=\frac{1}{\eta}\left[ t-\tau\left( 1-e^{-\nicefrac{t}{\tau}} \right) \right]$} & { $h(t)=\frac{1}{\eta} \left( 1-e^{-\nicefrac{t}{\tau}} \right)$} & \thead[l]{{ $\psi(t)=\frac{1}{m_R} e^{-\nicefrac{t}{\tau}} $,} \\ { \quad \quad \quad \quad \quad \quad \quad \quad $\tau=\frac{m_R}{\eta}=\frac{m}{6\pi R \eta}$} } \\

\thead[l]{Kelvin--Voigt Solid with\\elasticity $G$ and viscosity $\eta$ \\ \includegraphics[scale=0.30]{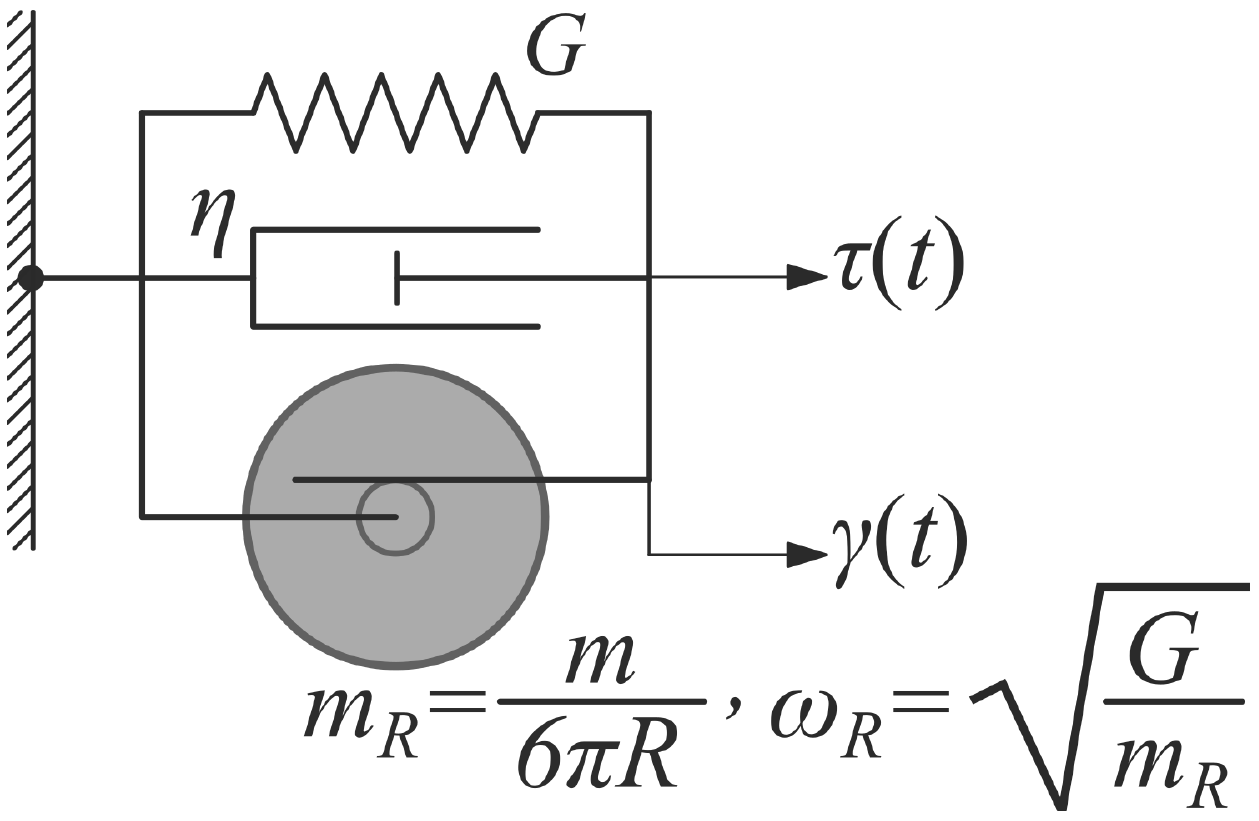}} & \thead[l]{ $J(t)=\frac{1}{m_R} \frac{1}{\omega_0^2} \Bigg[ 1-e^{-\nicefrac{t}{2\tau}} \Bigg( \cos (\omega_D t) $ \\  $ +\frac{1}{2\tau\omega_D} \sin (\omega_D t) \Bigg) \Bigg]$, \quad $\omega_D \tau>\frac{1}{2}$} & \thead[l]{ $h(t)=\frac{1}{m_R} \frac{1}{\omega_D} e^{-\nicefrac{t}{2\tau}} \sin (\omega_D t)$ \\ \\  $\omega_D=\omega_0\sqrt{1-\left(\frac{1}{2\omega_0 \tau}\right)^2}$, $\omega_D \tau>\frac{1}{2}$} & \thead[l]{ $\psi(t)=\frac{1}{m_R}e^{-\nicefrac{t}{2\tau}}\Bigg[\cos (\omega_D t)$ \\ \quad \quad \quad $- \frac{1}{2} \frac{1}{\tau \omega_D} \sin(\omega_D t)  \Bigg] $, \quad $\omega_D \tau>\frac{1}{2}$} \\

\thead[l]{Maxwell Fluid with elasticity\\$G$ and viscosity $\eta$ \\ \includegraphics[scale=0.30]{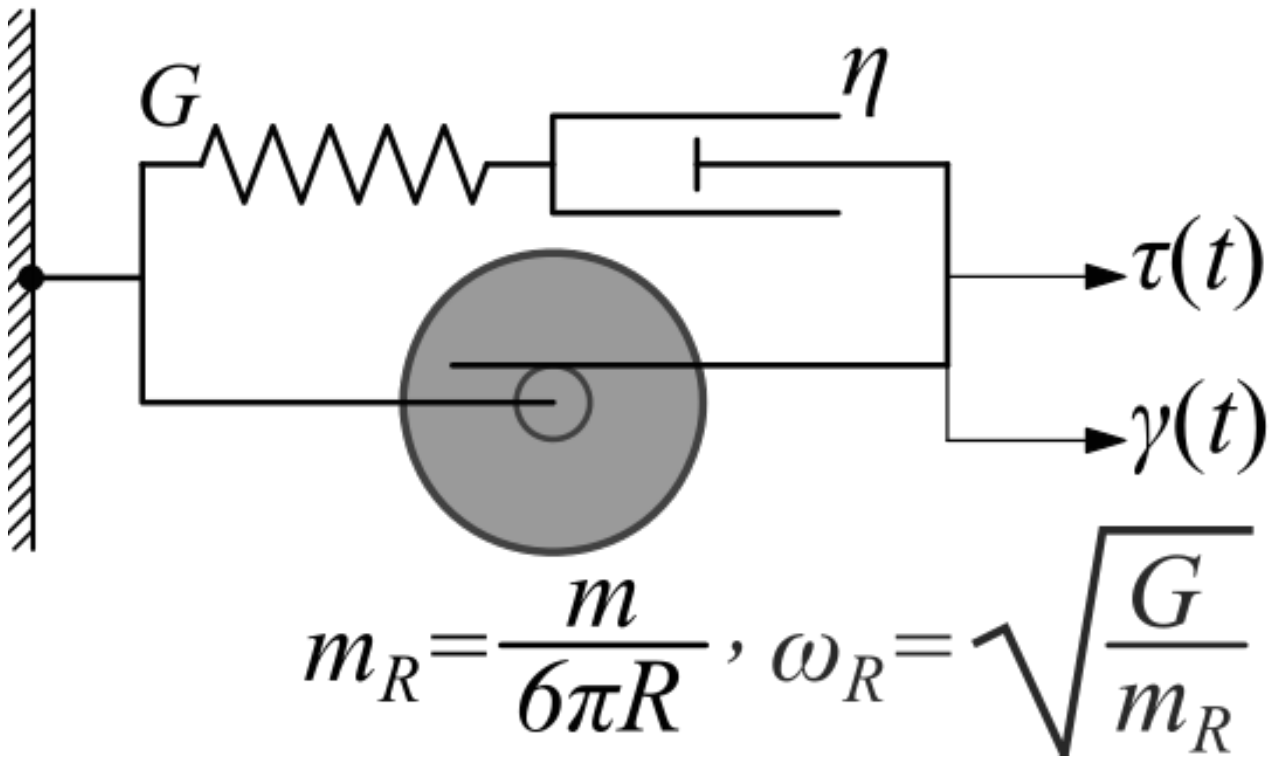}} & \thead[l]{ $J(t)=\frac{m_R}{\eta^2} \Bigg\lbrace \frac{t}{\tau} - \frac{1}{4\beta^2} \Bigg[ 4\beta^2 -1 -e^{-2\frac{t}{\tau}\beta^2} $ \\ $\times\Bigg( \frac{4\beta^3 -3 \beta}{\sqrt{1-\beta^2}}\sin\left( 2\frac{t}{\tau}\beta \sqrt{1-\beta^2} \right) +\left( 4\beta^2 -1 \right) $ \\ $\times\cos\left( 2\frac{t}{\tau}\beta \sqrt{1-\beta^2} \right) \Bigg) \Bigg] \Bigg\rbrace$, \enskip $\beta=\frac{\tau\omega_R}{2}<1$} & \thead[l]{ $h(t)=\frac{1}{\eta} \Bigg[ U(t-0) - e^{-2\frac{t}{\tau}\beta^2}$ \\ $\times \Bigg( \frac{2\beta^2-1}{2\beta\sqrt{1-\beta^2}} \sin\left( 2\beta\sqrt{1-\beta^2}\frac{t}{\tau} \right)$ \\ $ + \cos\left( 2\beta\sqrt{1-\beta^2}\frac{t}{\tau} \right) \Bigg) \Bigg]$, $\beta=\frac{\tau\omega_R}{2}<1$} & \thead[l]{ $\psi(t)=\frac{1}{m_R}e^{-2\frac{t}{\tau}\beta^2} \Bigg[ \cos\left( 2\frac{t}{\tau}\beta \sqrt{1-\beta^2} \right)$ \\ $+\frac{\beta}{\sqrt{1-\beta^2}} \sin\left( 2\frac{t}{\tau}\beta \sqrt{\beta^2-1}\right) \Bigg]$, $\beta=\frac{\tau\omega_R}{2}<1$}  \\
	
\thead[l]{Scott-Blair subdiffusive fluid\\with material constant $\mu_{\alpha}$\\ \includegraphics[scale=0.30]{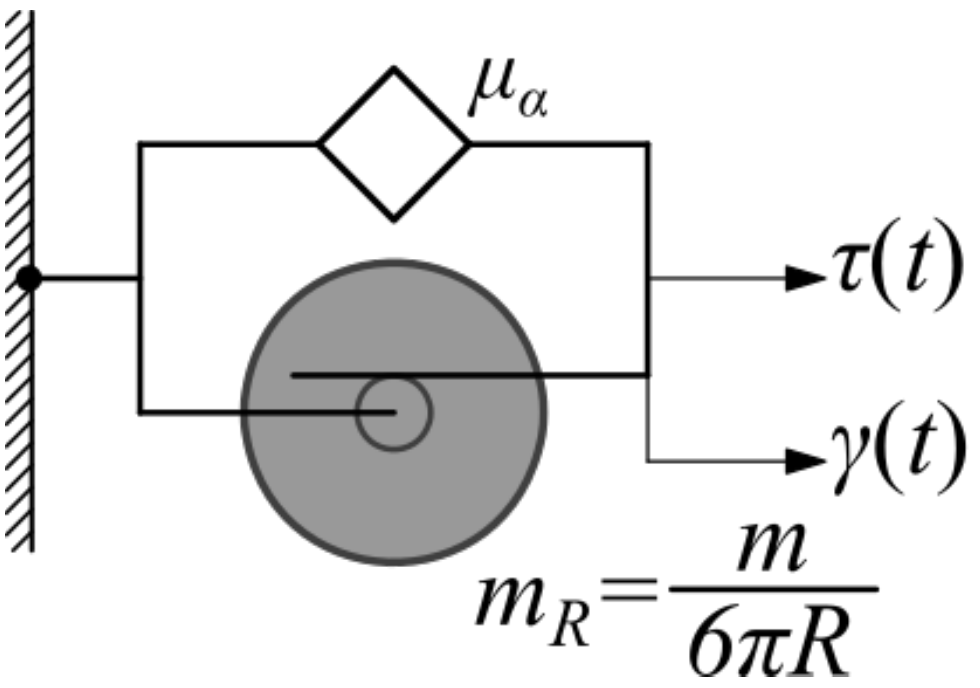}} & { $J(t)= \frac{1}{m_R}t^2 E_{2-\alpha ,\, 3}\left( -\frac{\mu_{\alpha}}{m_R}t^{2-\alpha} \right)$} & { $h(t)= \frac{1}{m_R}t E_{2-\alpha ,\, 2}\left( -\frac{\mu_{\alpha}}{m_R}t^{2-\alpha} \right)$} & { $\psi(t)= \frac{1}{m_R} E_{2-\alpha ,\, 1}\left( -\frac{\mu_{\alpha}}{m_R}t^{2-\alpha} \right)$} \\
\br
\end{tabular*}
\label{tab:Table1}
\end{sidewaystable}
\begin{eqnarray}\label{eq:Eq28}
\frac{m \omega_0}{2 N K_B T}  \frac{\mathrm{d}\left\langle \Delta r^{2} (t) \right\rangle}{\mathrm{d}t} & = \sqrt{G m_R} h (t) \\ \nonumber 
& = \frac{1}{\sqrt{\left( \frac{1}{2 \omega_0 \tau} \right)^2 - 1}} e^{-\nicefrac{t}{2 \tau}} \sinh \left( \omega_0 \tau \sqrt{\left( \frac{1}{2 \omega_0 \tau} \right)^2 - 1} \frac{t}{\tau} \right)
\end{eqnarray}
For small values of the dimensionless product $\omega_0 \tau$ (weak spring), Eq. (\ref{eq:Eq28}) at early times contracts to the solution for Brownian motion of particles in a Newtonian viscous fluid since the inertia and viscous terms dominate over the elastic term
\begin{equation}\label{eq:Eq29}
\frac{m\omega_0}{2 N K_B T} \frac{\mathrm{d}\left\langle \Delta r^{2} (t) \right\rangle}{\mathrm{d}t} = \omega_0 \tau (1-e^{-\nicefrac{t}{\tau}}).
\end{equation}
\noindent Equation (\ref{eq:Eq29}) is obtained after multiplying both sides of Eq. (\ref{eq:Eq10}) with $\omega_0$ and replacing $\frac{1}{\eta}$ with $\frac{6 \pi R \tau}{m}$. Figure \ref{fig:Fig03} plots the normalized impulse response function given by Eqs. (\ref{eq:Eq27}) and (\ref{eq:Eq28}) as a function of the dimensionless time $\frac{t}{\tau}$ for various values of $\omega_0 \tau = \frac{\sqrt{k m}}{6 \pi R} = \frac{\sqrt{G m_R}}{\eta}$ together with the results from Eq. (\ref{eq:Eq29}) (Newtonian viscous fluid) for values of $\omega_0 \tau=$ 0.2 and 0.3. Figure \ref{fig:Fig03} indicates that at large times the impulse response function, $h(t)$ for Brownian motion in a solid-like material (Kelvin-Voigt solid) vanishes; therefore, for such solid-like materials the one-sided sine and cosine integral transforms introduced by Nishi \textit{et al.} \cite{NishiKilfoilSchmidtMacKintosh2018} converge.

The results reached in Sections \ref{sec:Sec02} and \ref{sec:Sec03} are summarized in Table \ref{tab:Table1} which lists the mean-square displacement together with its first and second time-derivatives (velocity autocorrelation function) of Brownian microparticles suspended in a Newtonian fluid, a Kelvin--Voigt solid, a Maxwell fluid and a subdiffusive Scott--Blair fluid. Table \ref{tab:Table1} also shows the rheological analogues for the Brownian motion of microparticles suspended in the above mentioned viscoelastic materials together with the expressions of the corresponding deterministic creep compliance $J(t)=\frac{3\pi R}{N K_B T}\left\langle \Delta r^{2} (t) \right\rangle$, impulse response function $h(t)=\frac{3\pi R}{N K_B T}\frac{\mathrm{d}\left\langle \Delta r^{2} (t) \right\rangle}{\mathrm{d}t}$ and impulse strain-rate response function $\psi(t)=\frac{3\pi R}{N K_B T}\frac{\mathrm{d}^2\left\langle \Delta r^{2} (t) \right\rangle}{\mathrm{d}t^2}=\frac{6\pi R}{N K_B T}\left\langle v(0)v(t) \right\rangle$ of the viscoelastic material--inerter parallel connection.

\section{Impulse Response Function for Brownian Motion in a Maxwell Fluid}\label{sec:Sec04}

\begin{figure}[t!]
\centering
\includegraphics[width=.6\linewidth, angle=0]{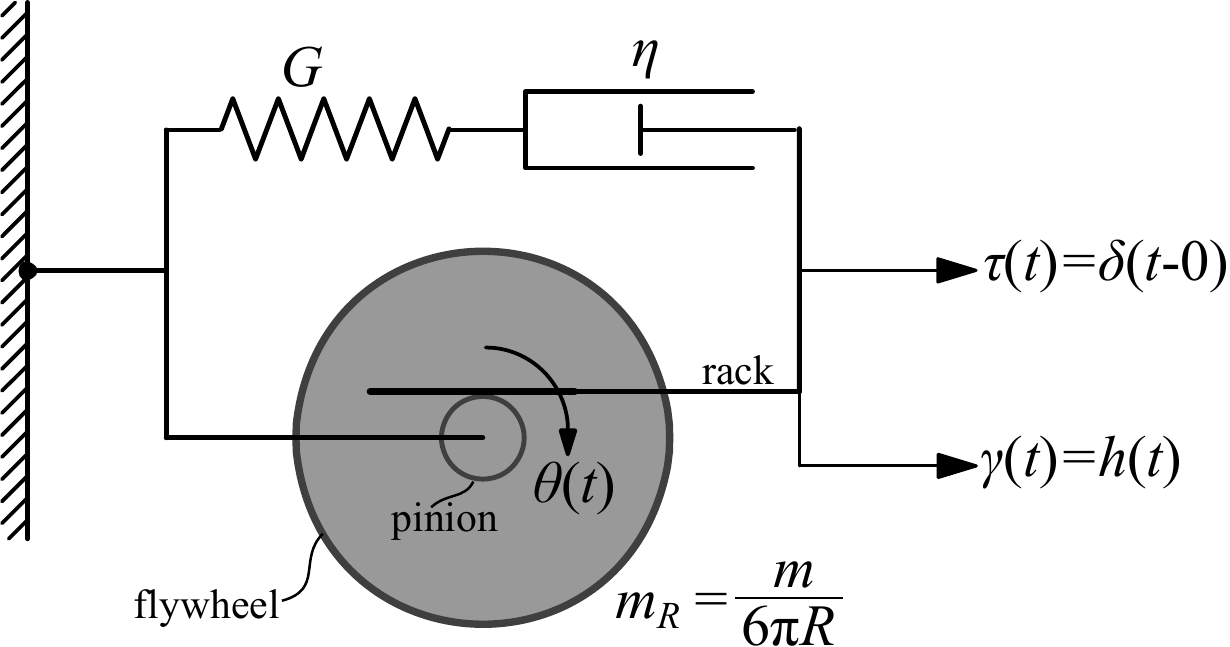}
\caption{Mechanical analogue for Brownian motion in a Maxwell fluid. It consists of the Maxwell element with shear modulus $G$ and shear viscosity $\eta$ that is connected in parallel with m an inerter with distributed inertance $m_R = \frac{m}{6\pi R}$.}
\label{fig:Fig04}
\end{figure}

\begin{figure}[t!]
\centering
\includegraphics[width=.7\linewidth, angle=0]{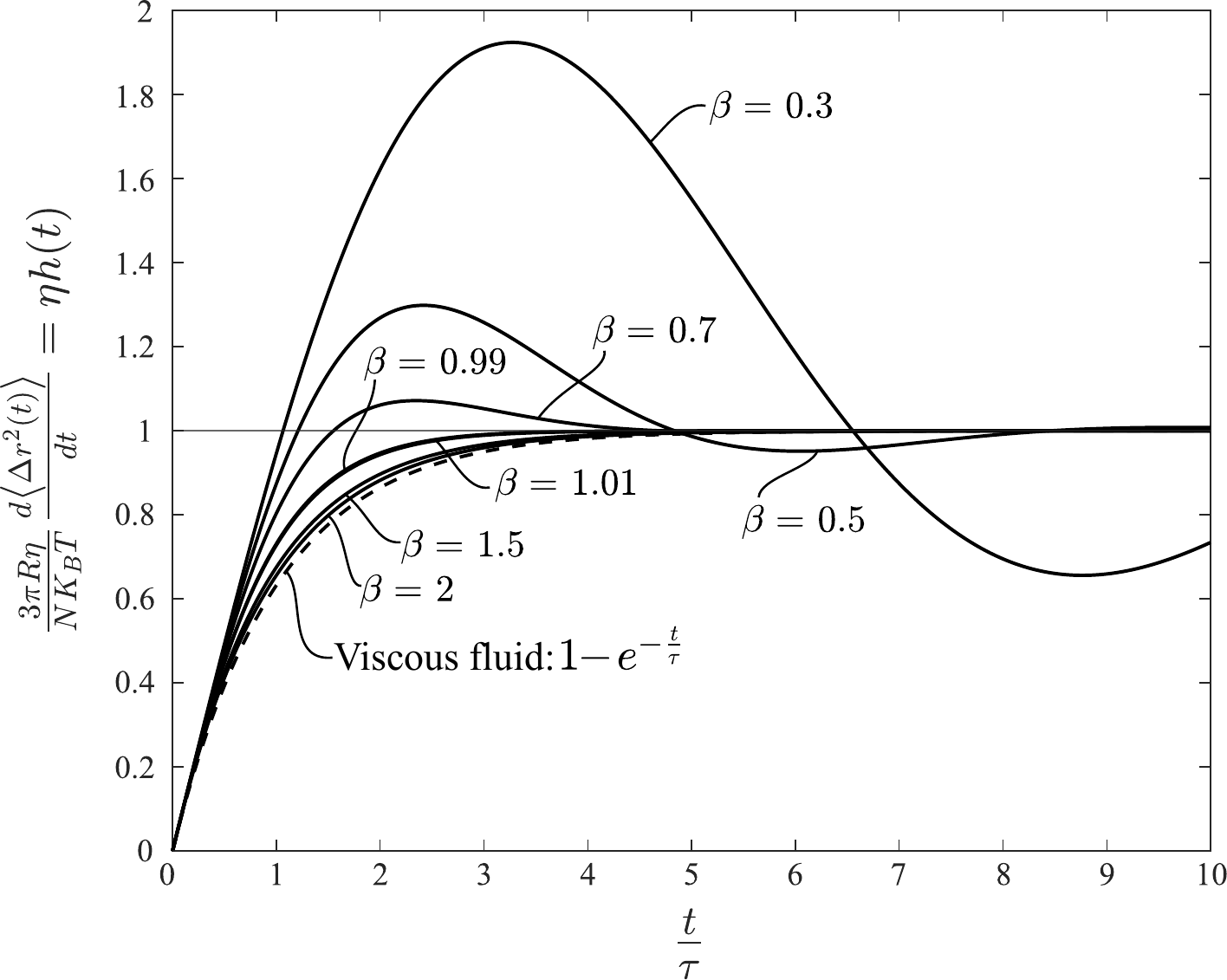}
\caption{Normalized time-derivative of the mean-square displacement of Brownian particles with mass $m$ and radius $R$ suspended in a Maxwell fluid with shear modulus $G$ and viscosity $\eta$ for various values of the parameter $\beta=\frac{1}{2\eta}\sqrt{\frac{mG}{6\pi R}}=\frac{1}{2\eta}\sqrt{m_R}\sqrt{G}$. For large values of $\beta$ (stiff spring), the solution contracts to the solution for Brownian motion of particles immersed in a Newtonian viscous fluid: $1-e^{-\nicefrac{t}{\tau}}$.}
\label{fig:Fig05}
\end{figure}

According to the viscous--viscoelastic correspondence principle for Brownian motion illustrated in Fig. \ref{fig:Fig01}, the rheological analogue for Brownian motion of particles suspended in a Maxwell fluid with a single relaxation time $\lambda=\frac{\eta}{G}$ is the mechanical network shown in Fig. \ref{fig:Fig04} which is a parallel connection of a Maxwell element with shear modulus $G$, and shear viscosity $\eta$, with an inerter with distributed inertance $m_R = \frac{m}{6\pi R}$. The mechanical network shown in Fig. \ref{fig:Fig04} is described by a third-order constitutive equation \cite{Makris2020}
\begin{equation}\label{eq:Eq30}
\tau(t)+\frac{\eta}{G} \frac{\mathrm{d}\tau(t)}{\mathrm{d}t}=\eta\frac{\mathrm{d}\gamma(t)}{\mathrm{d}t} + m_R\frac{\mathrm{d}^{2}\gamma(t)}{\mathrm{d}t^{2}} + \frac{\eta \, m_R}{G}\frac{\mathrm{d}^{3}\gamma(t)}{\mathrm{d}t^{3}}
\end{equation}
By defining the dissipation time $\tau=\frac{m_R}{\eta}=\frac{m}{6\pi R \eta}$ and the rotational angular frequency $\omega_R=\sqrt{\frac{G}{m_R}}=\sqrt{\frac{6\pi R G}{m}}$, Eq. (\ref{eq:Eq30}) assumes the form
\begin{equation}\label{eq:Eq31}
\tau(t) + \frac{1}{\tau \omega_R^{2}} \frac{\mathrm{d}\tau(t)}{\mathrm{d}t}  =  m_R \left( \frac{1}{\tau} \frac{\mathrm{d}\gamma(t)}{\mathrm{d}t} + \frac{\mathrm{d}^{2}\gamma(t)}{\mathrm{d}t^{2}} +\frac{1}{\tau \omega_R^{2}} \frac{\mathrm{d}^{3}\gamma(t)}{\mathrm{d}t^{3}} \right)
\end{equation}
The Laplace transform of Eq. (\ref{eq:Eq31}) gives $\gamma(s)=\mathcal{J}(s)\tau(s)$ where $\mathcal{J}(s)$ is the complex dynamic compliance of the mechanical network shown in Fig. \ref{fig:Fig04}
\begin{equation}\label{eq:Eq32}
\mathcal{J}(s)=\frac{1}{\mathcal{G}(s)}=\frac{\gamma(s)}{\tau(s)}=\frac{1}{m_R} \frac{1}{s} \frac{1+\frac{1}{\tau\omega_R^{2}}s}{\left( \frac{1}{\tau} + s + \frac{1}{\tau\omega_R^{2}}s^{2} \right)}
\end{equation}
In addition to $s=0$, the other two poles of the complex dynamic compliance $\mathcal{J}(s)$ given by Eq. (\ref{eq:Eq32}) are
\begin{equation}\label{eq:Eq33}
s_{1}=-\frac{\tau\omega_R^{2}}{2}+\omega_R\sqrt{\left( \frac{\tau\omega_R}{2} \right)^{2}-1}=-\omega_R\left( \beta-\sqrt{\beta^{2}-1} \right)
\end{equation}
and
\begin{equation}\label{eq:Eq34}
s_{2}=-\frac{\tau\omega_R^{2}}{2}-\omega_R\sqrt{\left( \frac{\tau\omega_R}{2} \right)^{2}-1}=-\omega_R\left( \beta+\sqrt{\beta^{2}-1} \right)
\end{equation}
where $\beta=\frac{\tau\omega_R}{2}=\frac{1}{2\eta}\sqrt{\frac{mG}{6\pi R}}$ is a dimensionless parameter of the mechanical network shown in Fig. \ref{fig:Fig04} and of the Brownian particle--Maxwell fluid system. By virtue of Eqs. (\ref{eq:Eq33}) and (\ref{eq:Eq34}), the complex dynamic compliance $\mathcal{J}(s)= \displaystyle\int_{0}^{\infty}h(t)e^{-st}\,\mathrm{d}t$ given by Eq. (\ref{eq:Eq32}) is expressed as
\begin{equation}\label{eq:Eq35}
\fl \mathcal{J}(s)=\frac{1}{\eta}  \Bigg[ \frac{1}{s}  - \left( \frac{2\beta^{2}-1}{4\beta\sqrt{\beta^{2}-1}} + \frac{1}{2}\right)\frac{1}{s-s_{1}} +  \left( \frac{2\beta^{2}-1}{4\beta\sqrt{\beta^{2}-1}} -\frac{1}{2}\right)\frac{1}{s-s_{2}} \Bigg]
\end{equation}
For the case where $\beta=\frac{\tau\omega_R}{2}>1$ (stiff spring) and by using that $\omega_R t=2\beta\frac{t}{\tau}$, the inverse Laplace transform of Eq. (\ref{eq:Eq35}) gives
\begin{eqnarray}\label{eq:Eq36}
h(t)=\mathcal{L}^{-1}\left\lbrace \mathcal{J}(s) \right\rbrace &= \frac{1}{\eta} \Bigg[ U(t-0) - e^{-2\frac{t}{\tau}\beta^2} \Bigg( \cosh\left( 2\beta\sqrt{\beta^2-1}\frac{t}{\tau} \right) \\ \nonumber 
& + \frac{2\beta^2-1}{2\beta\sqrt{\beta^2-1}}  \sinh\left( 2\beta\sqrt{\beta^2-1}\frac{t}{\tau} \right) \Bigg) \Bigg], \quad \beta>1
\end{eqnarray}
where $U(t-0)$ is the Heaviside unit-step function \cite{Lighthill1958}. For the case where $\beta=\frac{\tau\omega_R}{2}<1$ (flexible spring) the impulse response function of the mechanical network shown in Fig. \ref{fig:Fig04} is 
\begin{eqnarray}\label{eq:Eq37}
h(t)=\mathcal{L}^{-1}\left\lbrace \mathcal{J}(s) \right\rbrace & = \frac{1}{\eta} \Bigg[ U(t-0) - e^{-2\frac{t}{\tau}\beta^2} \Bigg(  \cos\left( 2\beta\sqrt{1-\beta^2}\frac{t}{\tau} \right) \\ \nonumber 
& + \frac{2\beta^2-1}{2\beta\sqrt{1-\beta^2}}   \sin\left( 2\beta\sqrt{1-\beta^2}\frac{t}{\tau} \right)  \Bigg) \Bigg], \quad \beta<1
\end{eqnarray}
Figure \ref{fig:Fig05} plots the normalized time-derivative of the mean-square displacement for Brownian motion in a Maxwell fluid
\begin{equation}\label{eq:Eq38}
\frac{3\pi R \eta}{N K_B T} \frac{\mathrm{d}\left\langle \Delta r^{2} (t) \right\rangle}{\mathrm{d}t} = \eta h(t)
\end{equation}
where the impulse response function, $h(t)$, is offered by Eq. (\ref{eq:Eq36}) or (\ref{eq:Eq37}) depending on the value of $\beta=\frac{\tau\omega_R}{2}$. For large values of $\beta$ (stiff spring) the solution contracts to the solution for Brownian motion of particles suspended in a Newtonian viscous fluid since the dashpot essentially reacts to a non-compliant element.

\section{Brownian Motion within a Viscoelastic Fluid Described by a Maxwell Element Connected in Parallel with a Dashpot}\label{sec:Sec05}

\begin{figure}[t!]
\centering
\includegraphics[width=.6\linewidth, angle=0]{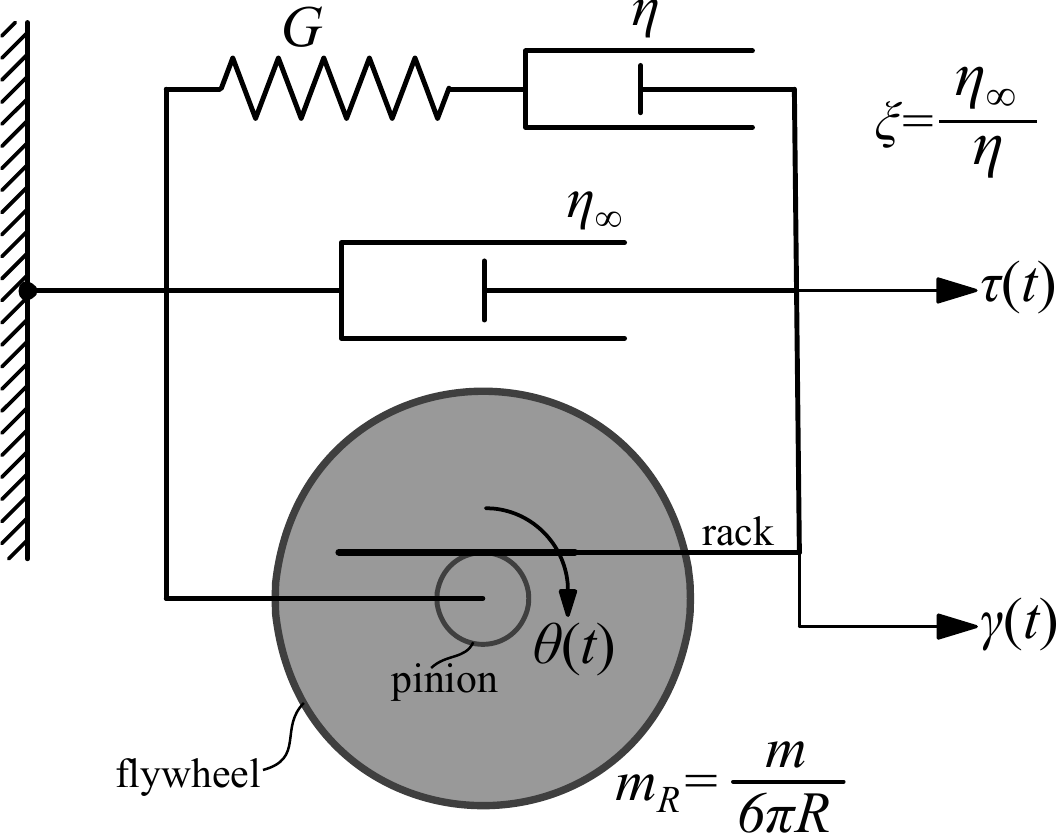}
\caption{Mechanical analogue for Brownian motion of microspheres with mass $m$ and radius $R$ immersed in a viscoelastic fluid described with a Maxwell element--dashpot parallel connection.}
\label{fig:Fig06new}
\end{figure}

The Maxwell element (a spring $G$ and a dashpot $\eta$ connected in series) when connected in parallel with a dashpot with viscosity $\eta_{\infty}$ may capture the linear response of selected soft materials such as wormlike micellar solutions and concentrated dispersions \cite{KhanMason2014,KhanMason2014PRE}. At very low frequencies there is a slow relaxation typically arising from the reorganizations of the colloidal structure in the viscoelastic material with relaxation time $\lambda=\frac{\eta}{G}$. At high frequencies because of the compliant spring $G$, the shear stresses are primarily resisted by the parallel dashpot with viscosity $\eta_{\infty}$, and the response is viscously dominated. Accordingly, the relaxation modulus, $G_{\textit{ve}}(t)$ of the Maxwell element--dashpot parallel connection is
\begin{equation}\label{eq:Eq39new}
G_{\textit{ve}}(t)=\eta_{\infty}\delta(t-0)+Ge^{-\frac{G}{\eta}t}
\end{equation}
In this section, the mean-square displacement of Brownian particles suspended in a Maxwell element--dashpot parallel connection is calculated with the correspondence principle summarized in Fig. \ref{fig:Fig01}. Accordingly, the problem reduces to the calculation of the creep compliance of a Maxwell element with shear modulus $G$ and shear viscosity $\eta$ that is connected in parallel with a dashpot with shear viscosity $\eta_{\infty}$ and the entire viscoelastic fluid is connected in parallel with an inerter with distributed inertance $m_R=\frac{m}{6\pi R}$ as shown in Fig. \ref{fig:Fig06new}. 

The total stress $\tau(t)=\tau_1(t)+\tau_2(t)+\tau_3(t)$ from the linear network shown in Fig. \ref{fig:Fig06new} is the summation of the stress output from the Maxwell element, $\tau_1(t)$
\begin{equation}\label{eq:Eq40new}
\tau_1(t)+\frac{\eta}{G}\frac{\mathrm{d}\tau_1(t)}{\mathrm{d}t}=\eta\frac{\mathrm{d}\gamma(t)}{\mathrm{d}t},
\end{equation}
the stress output from the dashpot with viscosity $\eta_{\infty}$, $\tau_2(t)$
\begin{equation}\label{eq:Eq41new}
\tau_2(t)=\eta_{\infty}\frac{\mathrm{d}\gamma(t)}{\mathrm{d}t}
\end{equation}
and the stress output from the inerter with distributed inertance $m_R$, $\tau_3(t)$
\begin{equation}\label{eq:Eq42new}
\tau_3(t)=m_R\frac{\mathrm{d}^2\gamma(t)}{\mathrm{d}t^2}
\end{equation}
The summation of Eqs. (\ref{eq:Eq40new}), (\ref{eq:Eq41new}) and (\ref{eq:Eq42new}) together with the time-derivatives of Eqs. (\ref{eq:Eq41new}) and (\ref{eq:Eq42new}) yields a third-order constitutive equation for the linear network shown in Fig. \ref{fig:Fig06new}
\begin{equation}\label{eq:Eq43new}
\fl  \tau(t)+\frac{\eta}{G}\frac{\mathrm{d}\tau(t)}{\mathrm{d}t} =m_R\left[ \frac{\eta}{m_R}\left( 1+\frac{\eta_{\infty}}{\eta} \right)\frac{\mathrm{d}\gamma(t)}{\mathrm{d}t} + \left( 1 + \frac{\eta \eta_{\infty}}{m_R G} \right)\frac{\mathrm{d^2}\gamma(t)}{\mathrm{d}t^2} + \frac{\eta}{G} \frac{\mathrm{d^3}\gamma(t)}{\mathrm{d}t^3}\right]
\end{equation}
By defining the dissipation time $\tau=\frac{m_R}{\eta}=\frac{m}{6 \pi R \eta}$, the rotational angular frequency, $\omega_R=\sqrt{\frac{G}{m_R}}=\sqrt{\frac{6 \pi R G}{m}}$ and the dimensionless viscosity ratio $\xi=\frac{\eta_{\infty}}{\eta}$, Eq. (\ref{eq:Eq43new}) assumes the form
\begin{equation}\label{eq:Eq44new}
\fl \tau(t)+\frac{1}{\tau \omega_R^2} \frac{\mathrm{d}\tau(t)}{\mathrm{d}t} = m_R \left[ \frac{1}{\tau} (1+\xi) \frac{\mathrm{d}\gamma(t)}{\mathrm{d}t} + \left( 1+\frac{\xi}{\tau^2 \omega_R^2} \right)\frac{\mathrm{d^2}\gamma(t)}{\mathrm{d}t^2} + \frac{1}{\tau \omega_R^2} \frac{\mathrm{d^3}\gamma(t)}{\mathrm{d}t^3} \right]
\end{equation}
Equation (\ref{eq:Eq44new}) is of the same form as Eq. (\ref{eq:Eq31}); however now the coefficients of the first and second time-derivatives of the shear strain contain the viscosity ratio $\xi=\frac{\eta_{\infty}}{\eta}$ which controls the effects of the in-parallel dashpot that its viscosity $\eta_{\infty}$ becomes dominant at high frequencies. The Laplace transform of Eq. (\ref{eq:Eq44new}) gives $\gamma(s)=\mathcal{J}(s)\tau(s)$ where $\mathcal{J}(s)$ is the complex dynamic compliance of the linear network shown in Fig. \ref{fig:Fig06new}.
\begin{equation}\label{eq:Eq45new}
\mathcal{J}(s)=\frac{1}{\mathcal{G}(s)}=\frac{\gamma(s)}{\tau(s)}=\frac{1}{m_R}\frac{1+\frac{1}{\tau\omega_R^2}s}{s\left( \frac{1+\xi}{\tau} +\left( 1 +\frac{\xi}{\tau^2\omega_R^2} \right)s + \frac{1}{\tau\omega_R^2}s^2 \right)}
\end{equation}
In addition to $s=$ 0, the other two poles of the complex dynamic complaince $\mathcal{J}(s)$ given by Eq. (\ref{eq:Eq45new}) are
\begin{equation}\label{eq:Eq46new}
s_1=-q\frac{\tau\omega_R^2}{2}+\omega_R p \sqrt{\left(\frac{q\tau\omega_R}{2 p}\right)^2 -1}=-\omega_R \left( q\beta -p\sqrt{\left( \frac{q\beta}{p} \right)^2 -1} \right)
\end{equation}
and
\begin{equation}\label{eq:Eq47new}
s_2=-q\frac{\tau\omega_R^2}{2}-\omega_R p \sqrt{\left(\frac{q\tau\omega_R}{2 p}\right)^2 -1}=-\omega_R \left( q\beta +p\sqrt{\left( \frac{q\beta}{p} \right)^2 -1} \right)
\end{equation}
where $\beta=\frac{\tau\omega_R}{2}$ as defined earlier, $p=\sqrt{1+\xi}$ and $q=1+\frac{\xi}{\tau^2\omega_R^2}=1+\frac{\xi}{4 \beta^2}$ are parameters of the system that depend on the viscosity ratio $\xi=\frac{\eta_{\infty}}{\eta}$. By virtue of Eqs. (\ref{eq:Eq46new}) and (\ref{eq:Eq47new}), the complex dynamic compliance $\mathcal{J}(s)=\displaystyle\int_0^\infty h(t)e^{-st}\, \mathrm{d}t$ given by Eq. (\ref{eq:Eq45new}) is expressed as
{\small
\begin{equation}\label{eq:Eq48new}
\fl \mathcal{J}(s)=\frac{1}{\eta p^2}\left[ \frac{1}{s}- \left( \frac{2 q\beta^2-p^2}{4\beta p \sqrt{\left(\frac{q\beta}{p}\right)^2-1}} +\frac{1}{2}\right) \frac{1}{s-s_1}   + \left( \frac{2 q\beta^2-p^2}{4\beta p \sqrt{\left(\frac{q\beta}{p}\right)^2-1}} -\frac{1}{2}\right) \frac{1}{s-s_2}  \right]
\end{equation}}
For the radical $\sqrt{\left(\frac{q\beta}{p}\right)^2-1}$ appearing in Eqs. (\ref{eq:Eq46new}) and (\ref{eq:Eq47new}) to be real, $\frac{q\beta}{p}=\left( \beta + \frac{\xi}{4\beta} \right) \frac{1}{\sqrt{1+\xi}}>1$ which leads to the condition $\beta>\beta_{\textit{cr}}=\frac{\sqrt{1+\xi}+1}{2}$. 

\begin{figure}[t!]
\centering
\includegraphics[width=.7\linewidth, angle=0]{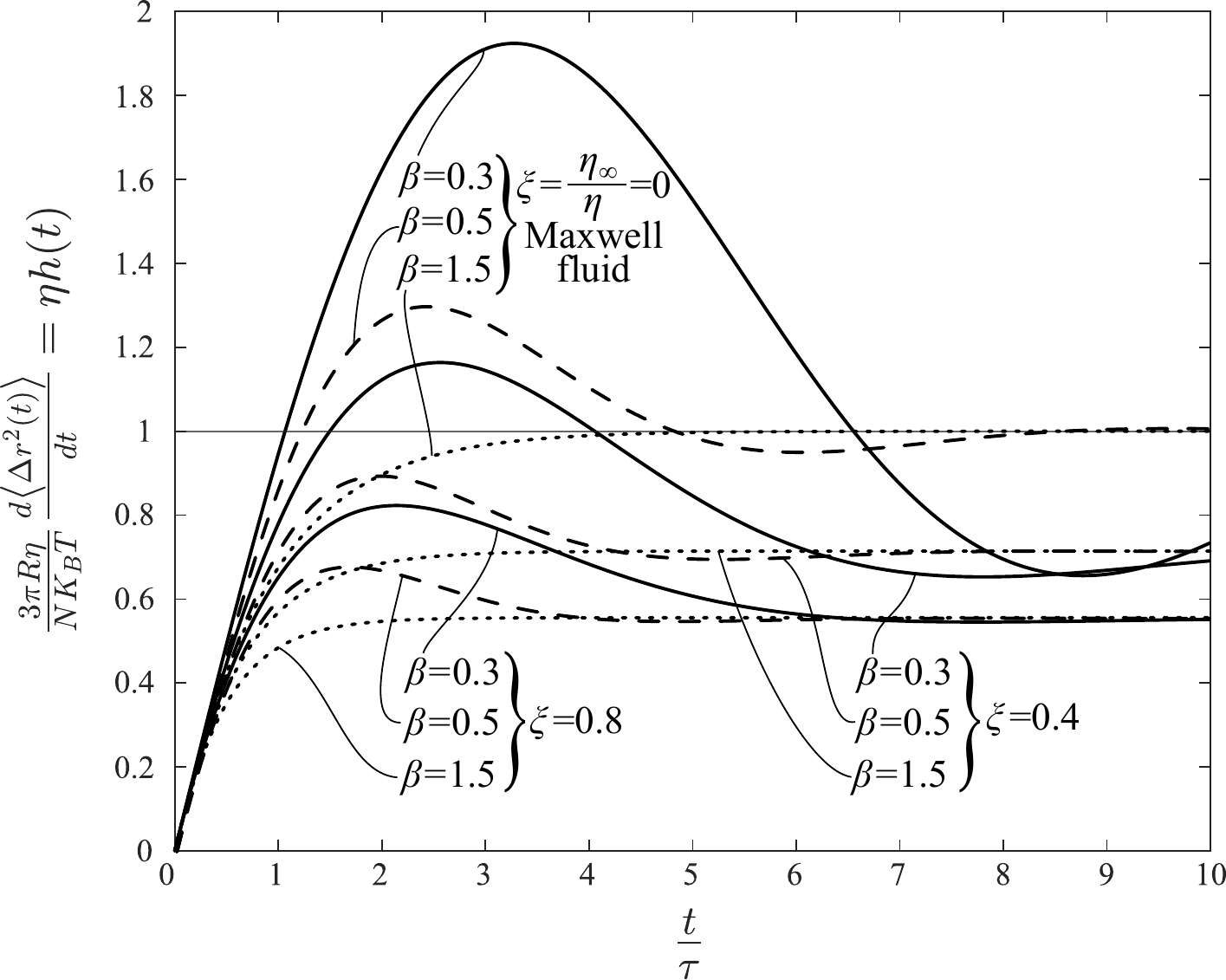}
\caption{Normalized time-derivative of the mean-square displacement of Brownian particles with mass $m$ and radius $R$ suspended in a viscoelastic fluid described by a Maxwell element--dashpot parallel connection. As the viscosity, $\eta_\infty$, of the parallel dashpot tends to zero, the response curves tend to those shown in Fig. \ref{fig:Fig05} which are for Brownian particles immersed in a Maxwell fluid ($\xi=0$). The impulse response function curves tend asymptotically to $\frac{1}{1+\xi}$.}
\label{fig:Fig07new}
\end{figure}

For the case where $\beta=\frac{\tau\omega_R}{2}>\beta_{\textit{cr}}=\frac{\sqrt{1+\xi}+1}{2}$ (stiff spring) and by using that $\omega_R t=2\beta\frac{t}{\tau}$, the inverse Laplace transform of Eq. (\ref{eq:Eq48new}) gives
\begin{eqnarray}\label{eq:Eq49new}
\fl h(t)=\mathcal{L}^{-1}\left\lbrace \mathcal{J}(s) \right\rbrace &= \frac{1}{\eta p^2} \left[\rule{0cm}{0.9cm}\right. U(t-0)-e^{-2 q \beta^2 \frac{t}{\tau}} \left(\rule{0cm}{0.9cm}\right. \cosh \left(2 \frac{t}{\tau}  \beta p  \sqrt{\left( \frac{q \beta}{p} \right)^2-1} \right) \\ \nonumber 
& + \frac{2 q\beta^2-p^2}{2\beta p \sqrt{\left( \frac{q \beta}{p} \right)^2-1}}   \sinh \left( 2\frac{t}{\tau}  \beta p  \sqrt{\left( \frac{q \beta}{p} \right)^2-1} \right)   \left.\rule{0cm}{0.9cm}\right) \left.\rule{0cm}{0.9cm}\right]
\end{eqnarray}
where $U(t-0)$ is the Heaviside unit-step function \cite{Lighthill1958}.
\\ \\
For the case where $\beta=\frac{\tau\omega_R}{2}<\beta_{\textit{cr}}=\frac{\sqrt{1+\xi}+1}{2}$ (flexible spring) the impulse response function of the mechanical network shown in Fig. \ref{fig:Fig06new} is
 \begin{eqnarray}\label{eq:Eq50new}
\fl h(t)=\mathcal{L}^{-1}\left\lbrace \mathcal{J}(s) \right\rbrace &= \frac{1}{\eta p^2} \left[\rule{0cm}{0.9cm}\right. U(t-0)-e^{-2 q \beta^2 \frac{t}{\tau}} \left(\rule{0cm}{0.9cm}\right. \cos \left(2 \frac{t}{\tau}  \beta p  \sqrt{1-\left( \frac{q \beta}{p} \right)^2} \right) \\ \nonumber 
& + \frac{2 q\beta^2-p^2}{2\beta p \sqrt{1-\left( \frac{q \beta}{p} \right)^2}}   \sin \left( 2\frac{t}{\tau}  \beta p  \sqrt{1-\left( \frac{q \beta}{p} \right)^2} \right)   \left.\rule{0cm}{0.9cm}\right) \left.\rule{0cm}{0.9cm}\right]
\end{eqnarray}
In the absence of the parallel dashpot shown in Fig. \ref{fig:Fig06new} ($\eta_\infty=\xi=0$), parameter $p=q=1$ and Eqs. (\ref{eq:Eq49new}) and (\ref{eq:Eq50new}) reduce to Eqs. (\ref{eq:Eq36}) and (\ref{eq:Eq37}). Figure \ref{fig:Fig07new} plots the normalized time-derivative of the mean-square displacement as expressed by Eq. (\ref{eq:Eq38}) for Brownian motion in a viscoelastic fluid described by a Maxwell element--dashpot parallel connection where the impulse response function, $h(t)$, is offered by Eq. (\ref{eq:Eq49new}) or (\ref{eq:Eq50new}) depending on the value of $\beta=\frac{\tau\omega_R}{2}$. The normalized impulse response curves shown in Fig. \ref{fig:Fig07new} tend asymptotically to $\frac{1}{1+\xi}$. Accordingly, at large times the impulse response function of the mechanical network shown in Fig. \ref{fig:Fig06new} is $h(t)=\frac{1}{\eta}\frac{1}{1+\xi}=\frac{1}{\eta+\eta_\infty}$ which is the correct long-term limit \cite{Makris2017}. Given this non-zero long-term value for this class of fluid-like viscoelastic materials \cite{KhanMason2014}, the sine and cosine integral transforms proposed by Nishi \textit{et al.} become ill-defined as was pointed out in the original paper \cite{NishiKilfoilSchmidtMacKintosh2018}.

According to the viscous--viscoelastic correspondence principle from Brownian motion illustrated in Fig. \ref{fig:Fig01}, the mean-square displacement of Brownian micro-spheres with mass $m$ and radius $R$ immersed in a viscoelastic fluid that is described by a Maxwell element connected in parallel with a dashpot is
\begin{equation}\label{eq:Eq51new}
\left\langle \Delta r^2 (t) \right\rangle = \frac{N K_B T}{3\pi R } J(t)
\end{equation}
where $J(t)$ is the creep compliance of the rheological network shown in Fig. \ref{fig:Fig06new}.

The Laplace transform of the creep compliance $J(t)$ is the complex creep function $\mathcal{C}(s)=\mathcal{L}\left\lbrace J(t) \right\rbrace = \displaystyle \int_0^\infty J(t) e^{-st}\, \mathrm{d}t=\frac{\mathcal{J}(s)}{s}$ \cite{PalmerXuWirtz1998,EvansTassieriAuhlWaigh2009, Makris2019,MakrisEfthymiou2020}. Accordingly, from Eq. (\ref{eq:Eq48new}) the complex creep function of the rheological network shown in Fig. \ref{fig:Fig06new} is
\begin{eqnarray}\label{eq:Eq52new}
\mathcal{C}(s)=\frac{\mathcal{J}(s)}{s}&=\frac{1}{\eta p^2}\left[ \frac{1}{s^2}- \left( \frac{2 q\beta^2-p^2}{4\beta p \sqrt{\left(\frac{q\beta}{p}\right)^2-1}} +\frac{1}{2}\right) \frac{1}{s(s-s_1)}  \right. \\ \nonumber
& \left. + \left( \frac{2 q\beta^2-p^2}{4\beta p \sqrt{\left(\frac{q\beta}{p}\right)^2-1}} -\frac{1}{2}\right) \frac{1}{s(s-s_2)}  \right]
\end{eqnarray}
\begin{figure}[t!]
\centering
\includegraphics[width=.7\linewidth, angle=0]{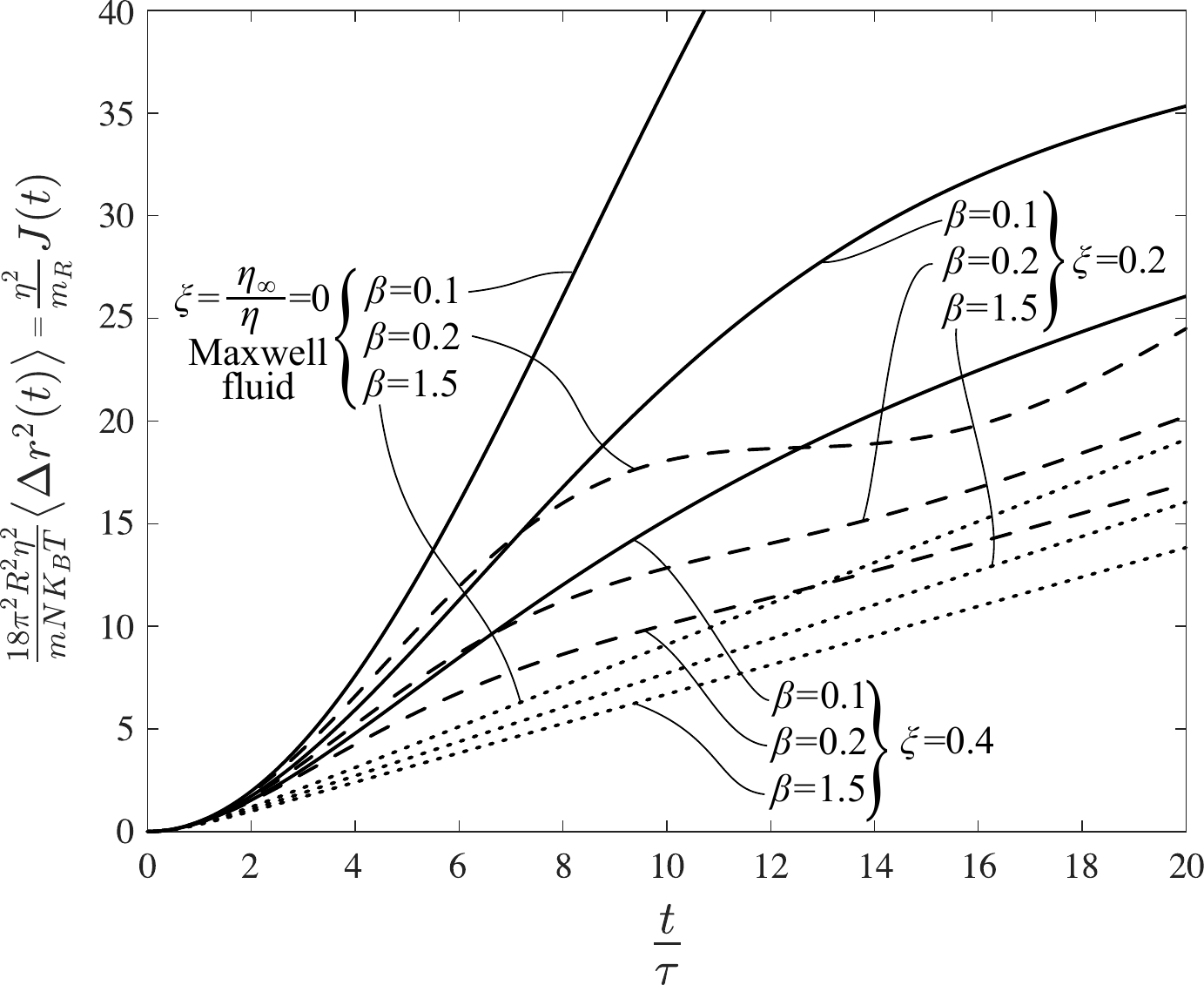}
\caption{Normalized mean-square displacement of Brownian microspheres suspended in a Maxwell element with shear modulus $G$ and shear viscosity $\eta$ connected in parallel with a dashpot with viscosity $\eta_\infty$.}
\label{fig:Fig08new}
\end{figure}
Using that $\mathcal{L}^{-1}\left\lbrace \frac{1}{s(s-s_j)} \right\rbrace = \frac{1}{s_j}\left( e^{s_jt}-1 \right)$ in association with that the poles $s_1$ and $s_2$ are given by Eqs. (\ref{eq:Eq46new}) and (\ref{eq:Eq47new}), the inverse Laplace transform of Eq. (\ref{eq:Eq52new}) for $\beta=\frac{\tau\omega_R}{2}>\beta_{\textit{cr}}=\frac{\sqrt{1+\xi}+1}{2}$ (stiff spring) is
\begin{eqnarray}\label{eq:Eq53new}
\fl J(t) =\mathcal{L}^{-1}\left\lbrace \mathcal{C}(s) \right\rbrace=\frac{m_R}{\eta^2 p^2} \left(\rule{0cm}{0.9cm}\right. \frac{t}{\tau} -\frac{1}{4p^2 \beta^2}  \left\lbrace\rule{0cm}{0.9cm}\right. 4 q \beta^2 - p^2 - e^{-2 q \beta^2 \frac{t}{\tau}} \\ \nonumber 
\times \left[\rule{0cm}{0.9cm}\right. \frac{4 q^2 \beta^3-p^2(2+q)\beta}{p\sqrt{\left( \frac{q\beta}{p} \right)^2-1}} \sinh \left( 2\frac{t}{\tau}  \beta p  \sqrt{\left( \frac{q \beta}{p} \right)^2-1} \right) \\ \nonumber 
 + \left( 4 q\beta^2 -p^2 \right)\cosh \left(2 \frac{t}{\tau}  \beta p  \sqrt{\left( \frac{q \beta}{p} \right)^2-1} \right)  \left.\rule{0cm}{0.9cm}\right] \left.\rule{0cm}{0.9cm}\right\rbrace \left.\rule{0cm}{0.9cm}\right),   \frac{q\beta}{p}> 1 \textit{ or } \beta>\frac{\sqrt{1+\xi}+1}{2}
\end{eqnarray}
%
%
For the case where $\beta=\frac{\tau\omega_R}{2}<\beta_{\textit{cr}}=\frac{\sqrt{1+\xi}+1}{2}$ (flexible spring) the creep compliance of the rheological network shown in Fig. \ref{fig:Fig06new} is
\begin{eqnarray}\label{eq:Eq54new}
\fl J(t)=\mathcal{L}^{-1}\left\lbrace \mathcal{C}(s) \right\rbrace=\frac{m_R}{\eta^2 p^2} \left(\rule{0cm}{0.9cm}\right. \frac{t}{\tau} -\frac{1}{4 p^2 \beta^2}  \left\lbrace\rule{0cm}{0.9cm}\right. 4 q \beta^2 - p^2 - e^{-2 q \beta^2 \frac{t}{\tau}} \\ \nonumber 
 \times \left[\rule{0cm}{0.9cm}\right. \frac{4 q^2 \beta^3-p^2(2+q)\beta}{p\sqrt{1-\left( \frac{q\beta}{p} \right)^2}} \sin \left( 2\frac{t}{\tau}  \beta p  \sqrt{1-\left( \frac{q \beta}{p} \right)^2} \right) \\ \nonumber 
 + \left( 4 q\beta^2 -p^2 \right)\cos \left(2 \frac{t}{\tau}  \beta p  \sqrt{1-\left( \frac{q \beta}{p} \right)^2} \right)  \left.\rule{0cm}{0.9cm}\right] \left.\rule{0cm}{0.9cm}\right\rbrace \left.\rule{0cm}{0.9cm}\right),   \frac{q\beta}{p}< 1 \textit{ or } \beta<\frac{\sqrt{1+\xi}+1}{2}
\end{eqnarray}
In the absence of the parallel dashpot shown in Fig. \ref{fig:Fig06new} $(\eta_\infty=0)$, parameters $p=q=1$ and Eq. (\ref{eq:Eq54new}) reduces to the expression shown in the first column -- third row of Table \ref{tab:Table1}. The time derivatives of the creep compliances offered by Eqs. (\ref{eq:Eq53new}) and (\ref{eq:Eq54new}) are the impulse response functions offered by Eqs. (\ref{eq:Eq49new}) and (\ref{eq:Eq50new}). By employing the correspondence principle for Brownian motion the mean-square displacement of Brownian particles immersed in a viscoelastic fluid that is described by a Maxwell element connected in parallel with a dashpot is given by Eq. (\ref{eq:Eq51new}) where the creep compliance $J(t)$ is offered by Eq. (\ref{eq:Eq53new}) or (\ref{eq:Eq54new}). Figure \ref{fig:Fig08new} plots the normalized mean-square displacement 
\begin{equation}\label{eq:Eq55new}
\frac{18\pi^2 R^2 \eta^2}{m N K_B T} \left\langle \Delta r^2 (t) \right\rangle = \frac{\eta^2}{m_R} J(t)
\end{equation}
as a function of the dimensionless time $\frac{t}{\tau}$ for various values of $\beta=\frac{\tau \omega_R}{2}=\frac{1}{2\eta}\sqrt{\frac{mG}{6 \pi R}}=\frac{1}{2}\frac{\sqrt{m_R}\sqrt{G}}{\eta}$ and $\xi=\frac{\eta_\infty}{\eta}$. For values of $\beta=\frac{\tau \omega_R}{2}<1$, the shear modulus $G$ is weak therefore, the inertia effects are more pronounced. In this case the mean-square displacement shown in Fig. \ref{fig:Fig08new} exhibits a reversal of curvature as the dimensionless time $\frac{t}{\tau}$ increases. As the viscosity $\eta_\infty$ of the parallel dashpot increases, the tendency for a plateau formation (that happens for Brownian motion in a Maxwell fluid\cite{vanZantenRufener2000}) tends to vanish and the behavior is dominated by viscosity. 

Equation (\ref{eq:Eq03}) in association with the Laplace transform of the mean-square displacement given by Eq. (\ref{eq:Eq51new})
\begin{equation}\label{eq:Eq56new}
\mathcal{L}\left\lbrace \left\langle \Delta r^2 (t) \right\rangle \right\rbrace = \left\langle \Delta r^2 (s) \right\rangle = \frac{N K_B T}{3 \pi R} \frac{\mathcal{J}(s)}{s} = \frac{N K_B T}{3 \pi R} \frac{1}{s \mathcal{G}(s)}
\end{equation}
shows that the Laplace transform of the velocity autocorrelation function is proportional to the inverse of the complex dynamic viscosity $\eta(s)=\frac{\mathcal{G}(s)}{s}$
\begin{equation}\label{eq:Eq57new}
\left\langle v(0)v(t) \right\rangle = \frac{s^2}{2} \left\langle \Delta r^2 (s) \right\rangle = \frac{N K_B T}{6 \pi R} \frac{s}{\mathcal{G}(s)} =\frac{N K_B T}{6 \pi R} \frac{1}{\eta (s)}
\end{equation}
The inverse of the complex dynamic viscosity is known in rheology as the complex dynamic fluidity $\phi(s)=\frac{1}{\eta(s)}=\frac{\dt{\gamma}(s)}{\tau(s)}$ \cite{Giesekus1995,MakrisKampas2009,MakrisEfthymiou2020} and relates a strain-rate output to a stress input. Accordingly, Eq. (\ref{eq:Eq57new}) is expressed as 
\begin{equation}\label{eq:Eq58new}
\left\langle v(0)v(s) \right\rangle =\frac{N K_B T}{6 \pi R} \phi(s)
\end{equation}
In structural mechanics, the equivalent of the complex dynamic fluidity $\phi(s)$ at the velocity--force level is known as the mobility or mechanical admittance \cite{HarrisCrede1976,Makris1997b}.

For the Maxwell element--dashpot--inerter parallel connection shown in Fig. \ref{fig:Fig06new}, the complex dynamic fluidity $\phi(s)=\frac{s}{\mathcal{G}(s)}$ derives directly from Eq. (\ref{eq:Eq45new})
\begin{equation}\label{eq:Eq59new}
\phi(s)=\frac{s}{\mathcal{G}(s)}=\frac{1}{m_R}\, \frac{1+\frac{1}{\tau\omega_R^2}s}{\frac{1+\xi}{\tau}+\left( 1+\frac{\xi}{\tau^2\omega_R^2}  \right)s + \frac{1}{\tau\omega_R^2}s^2}
\end{equation}
where the poles of the denominator of Eq. (\ref{eq:Eq59new}) are offered by Eqs. (\ref{eq:Eq46new}) and (\ref{eq:Eq47new}). The inverse Laplace transform of the complex dynamic fluidity $\mathcal{L}^{-1}\left\lbrace \phi(s) \right\rbrace = \psi(t)$ is the impulse strain-rate response function defined as the resulting strain-rate output at time $t$ due to an impulsive stress input $\tau(t)=\delta(t-\xi)$ with $\xi<t$. Accordingly, the velocity autocorrelation function of Brownian particles immersed in any isotropic, linear viscoelastic material is proportional to the impulse strain-rate response function of the viscoelastic matetial--inerter parallel connection
\begin{equation}\label{eq:Eq60new}
\left\langle v(0)v(t) \right\rangle =\frac{N K_B T}{6 \pi R} \mathcal{L}^{-1}\left\lbrace \phi(s) \right\rbrace = \frac{N K_B T}{6 \pi R} \psi(t)
\end{equation}
\begin{figure}[t!]
\centering
\includegraphics[width=.7\linewidth, angle=0]{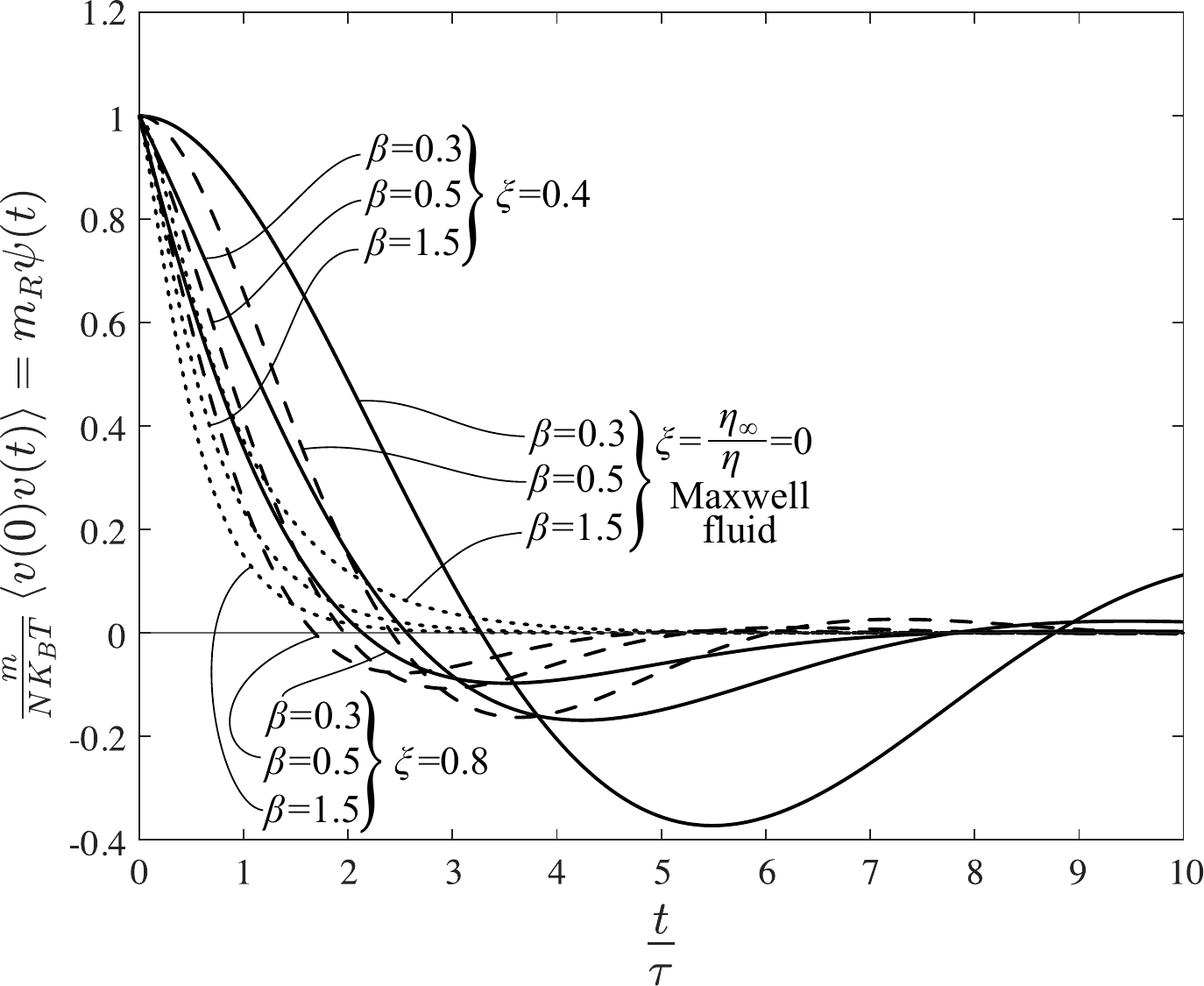}
\caption{Normalized velocity autocorrelation function of Brownian particles immersed in a viscoelastic material that is described by a Maxwell element connected in parallel with a dashpot with viscosity $\eta_\infty$.}
\label{fig:Fig09new}
\end{figure}
For the case where $\beta=\frac{\tau \omega_R}{2}>\beta_{\textit{cr}}=\frac{\sqrt{1+\xi}+1}{2}$ (stiff spring) the inverse Laplace transform of Eq. (\ref{eq:Eq59new}) gives
\begin{eqnarray}\label{eq:Eq61new}
\fl \psi(t)=\mathcal{L}^{-1}\left\lbrace \phi(s) \right\rbrace = \frac{1}{m_R} e^{-2 q \beta^2 \frac{t}{\tau}} \left[\rule{0cm}{0.9cm}\right. \cosh \left(2 \frac{t}{\tau}  \beta p  \sqrt{\left( \frac{q \beta}{p} \right)^2-1} \right) \\ \nonumber 
 + \frac{\beta(2-q)}{ p \sqrt{\left( \frac{q \beta}{p} \right)^2-1}}   \sinh \left( 2\frac{t}{\tau}  \beta p  \sqrt{\left( \frac{q \beta}{p} \right)^2-1} \right)   \left.\rule{0cm}{0.9cm}\right] ,   \beta=\frac{\tau \omega_R}{2}>\beta_{\textit{cr}}=\frac{\sqrt{1+\xi}+1}{2}
\end{eqnarray}
whereas, for the case where $\beta=\frac{\tau \omega_R}{2}<\beta_{\textit{cr}}=\frac{\sqrt{1+\xi}+1}{2}$ (flexible spring), the impulse strain-rate response function of the rheological network shown in Fig. \ref{fig:Fig06new} is
\begin{eqnarray}\label{eq:Eq62new}
\fl \psi(t)=\mathcal{L}^{-1}\left\lbrace \phi(s) \right\rbrace = \frac{1}{m_R} e^{-2 q \beta^2 \frac{t}{\tau}} \left[\rule{0cm}{0.9cm}\right. \cos \left(2 \frac{t}{\tau}  \beta p  \sqrt{1-\left( \frac{q \beta}{p} \right)^2} \right) \\ \nonumber 
 + \frac{\beta(2-q)}{ p \sqrt{1-\left( \frac{q \beta}{p} \right)^2}}   \sin \left( 2\frac{t}{\tau}  \beta p  \sqrt{1-\left( \frac{q \beta}{p} \right)^2} \right)   \left.\rule{0cm}{0.9cm}\right] ,  \beta=\frac{\tau \omega_R}{2}<\beta_{\textit{cr}}=\frac{\sqrt{1+\xi}+1}{2}
\end{eqnarray}
In the absence of the parallel dashpot shown in Fig. \ref{fig:Fig06new}, parameters $p=q=1$ and Eq. (\ref{eq:Eq62new}) reduces to the expression shown in the last column -- third row of Table \ref{tab:Table1}.

By virtue of Eq. (\ref{eq:Eq60new}), Fig. \ref{fig:Fig09new} plots the normalized velocity autocorrelation function for Brownian motion in a viscoelastic fluid that is described with a Maxwell element connected in parallel with a dashpot with viscosity $\eta_\infty$. In Eq. (\ref{eq:Eq60new}) the impulse strain-rate response function, $\psi(t)$ is offered by Eq. (\ref{eq:Eq61new}) or (\ref{eq:Eq62new}) depending on the value of $\beta=\frac{\tau\omega_R}{2}$.

\section{Impulse Response Function for Brownian Motion within a Subdiffusive Material}\label{sec:Sec06}

Several complex materials exhibit a subdiffusive behavior where from early times and over several temporal decades the mean-square displacement of suspended particles grows with time according to a power law; $\left\langle \Delta r^2 (t) \right\rangle \sim t^{\alpha}$, where 0 $\leq \alpha \leq$ 1 is the subdiffusive exponent \cite{PalmerXuWirtz1998,XuViasnoffWirtz1998,GislerWeitz1999,JeonChechkinMetzler2014,SafdariCherstvyChechkinBodrovaMetzler2017}. This type of a power-law rheological behavior was first reported by Nutting \cite{Nutting1921} who noticed that the stress response of several fluid-like materials when subjected to a step-strain $\left( \gamma (t) = U(t-0) \right)$ decays following a power law, $\tau(t)=G_{\textit{ve}}(t)\sim t^{-\alpha}$, where 0 $\leq \alpha \leq$ 1 and $G_{\textit{ve}}(t)$ is the relaxation modulus of the viscoelastic material. Following Nutting's \cite{Nutting1921} observations and the early work of Gemant \cite{Gemant1936,Gemant1938} on fractional differentials, Scott-Blair \cite{ScottBlair1944,ScottBlair1947} proposed the springpot element, which is a mechanical idealization in-between an elastic spring and a viscous dashpot with constitutive law
\begin{equation}\label{eq:Eq39}
\tau(t)=\mu_{\alpha} \frac{\mathrm{d}^{\alpha}\gamma(t)}{\mathrm{d}t^{\alpha}} , \quad 0 \leq \alpha \leq 1
\end{equation}
where $\alpha$ is a positive real number, 0 $\leq \alpha \leq$ 1, $\mu_{\alpha}$ is a phenomenological material parameter with units  $\left[M\right]\left[L\right]^{-1}\left[T\right]^{\alpha-2}$ (i.e. Pa s$^2$), and $\frac{\mathrm{d}^{\alpha}\gamma(t)}{\mathrm{d}t^{\alpha}}$ is the fractional derivative of order $\alpha$ of the strain history, $\gamma(t)$.

A definition of the fractional derivative of order $\alpha$ is given through the Riemann--Liouville convolution integral \cite{OldhamSpanier1974,SamkoKilbasMarichev1974,MillerRoss1993,Podlubny1998} 
\begin{equation}\label{eq:Eq40}
I^{\alpha}\gamma(t)=\frac{1}{\Gamma(\alpha)}\int_{0^-}^{t}\frac{1}{(t-\xi)^{1-\alpha}}\gamma(\xi)\, \mathrm{d}\xi , \quad \alpha \in \mathbb{R}^+
\end{equation}
where $\mathbb{R}^+$ is the set of positive real numbers and $\Gamma(\alpha)$ is the Gamma function. The integral in Eq. (\ref{eq:Eq40}) converges only for $\alpha>$ 1, or in the case where $\alpha$ is a complex number, the integral converges for $R(\alpha)>$ 0. Nevertheless, by a proper analytic continuation across the line $R(\alpha)=$ 0 and provided that the function $\gamma(t)$ is $n$-times differentiable, it can be shown that the integral given by Eq. (\ref{eq:Eq40}) exists for $n-R(\alpha)>$ 0 \cite{Riesz1949}. In this case, the fractional derivative of order $\alpha\in\mathbb{R}^+$ exists and is defined in the context of generalized functions as \cite{OldhamSpanier1974,Podlubny1998,Mainardi2010,Makris2021FF}
\begin{equation}\label{eq:Eq41}
\fl \frac{\mathrm{d}^{\alpha}\gamma(t)}{\mathrm{d}t^{\alpha}} =I^{-\alpha}\gamma(t) = \int_{0^-}^t \frac{\mathrm{d}^q\delta(t-\xi)}{\mathrm{d}t^q}\gamma(\xi)\, \mathrm{d}\xi  =\frac{1}{\Gamma(-\alpha)} \int_{0^-}^{t}\frac{1}{(t-\xi)^{1+\alpha}}\gamma(\xi)\, \mathrm{d}\xi , \quad \alpha \in \mathbb{R}^+
\end{equation}
where the lower limit of integration, 0$^-$ may engage an entire singular function at the time origin such as $\gamma(t)=\delta(t-0)$ \cite{Lighthill1958}.

The relaxation modulus $($stress history due to a unit-amplitude step-strain, $\gamma(t)=U(t-0))$ of the springpot element (Scott-Blair fluid) expressed by Eq. (\ref{eq:Eq39}) is \cite{SmitDeVries1970, Koeller1984, Friedrich1991, HeymansBauwens1994, SchiesselMetzlerBlumenNonnenmacher1995, PaladeVerneyAttane1996}
\begin{equation}\label{eq:Eq42}
G_{\textit{ve}}(t)=\mu_{\alpha}\frac{1}{\Gamma(1-\alpha)}t^{-\alpha} , \quad t>0
\end{equation}
which decays by following the power law initially observed by \cite{Nutting1921}. The creep compliance (retardation function) of the springpot element is \cite{Koeller1984, Friedrich1991, HeymansBauwens1994, SchiesselMetzlerBlumenNonnenmacher1995}
\begin{equation}\label{eq:Eq43}
J_{\textit{ve}}(t)=\frac{1}{\mu_{\alpha}}\frac{1}{\Gamma(1+\alpha)}t^{\alpha} , \quad t \geq 0
\end{equation}
The power law, $t^{\alpha}$, appearing in Eq. (\ref{eq:Eq43}) renders the elementary springpot element expressed by Eq. (\ref{eq:Eq39}) (Scott-Blair fluid), a suitable phenomenological model to study Brownian motion in subdiffusive materials.

The mean-square displacement of Brownian particles suspended in the fractional Scott-Blair fluid described by Eq. (\ref{eq:Eq39}) was evaluated in \cite{KobelevRomanov2000, Lutz2001} after computing the velocity autocorrelation function of the random motion of the suspended microspheres with $m$ and radius $R$,
\begin{equation}\label{eq:Eq44}
\left\langle \Delta r^{2} (t) \right\rangle = \frac{2 N K_B T}{m} t^2 E_{2-\alpha, \, 3}\left( -\frac{6\pi R \mu_{\alpha}}{m} t^{2-\alpha}\right)
\end{equation}
where $E_{\alpha, \, \beta}(z)$ is the two-parameter Mittag--Leffler function \cite{Erdelyi1953, GorenfloKilbasMainardiRogosin2014}
\begin{equation}\label{eq:Eq45}
E_{\alpha,  \, \beta}(z)=\sum_{j=0}^{\infty}\frac{z^j}{\Gamma(j\alpha+\beta)} , \quad \alpha,  \beta > 0
\end{equation}
\begin{figure}[t!]
\centering
\includegraphics[width=.6\linewidth, angle=0]{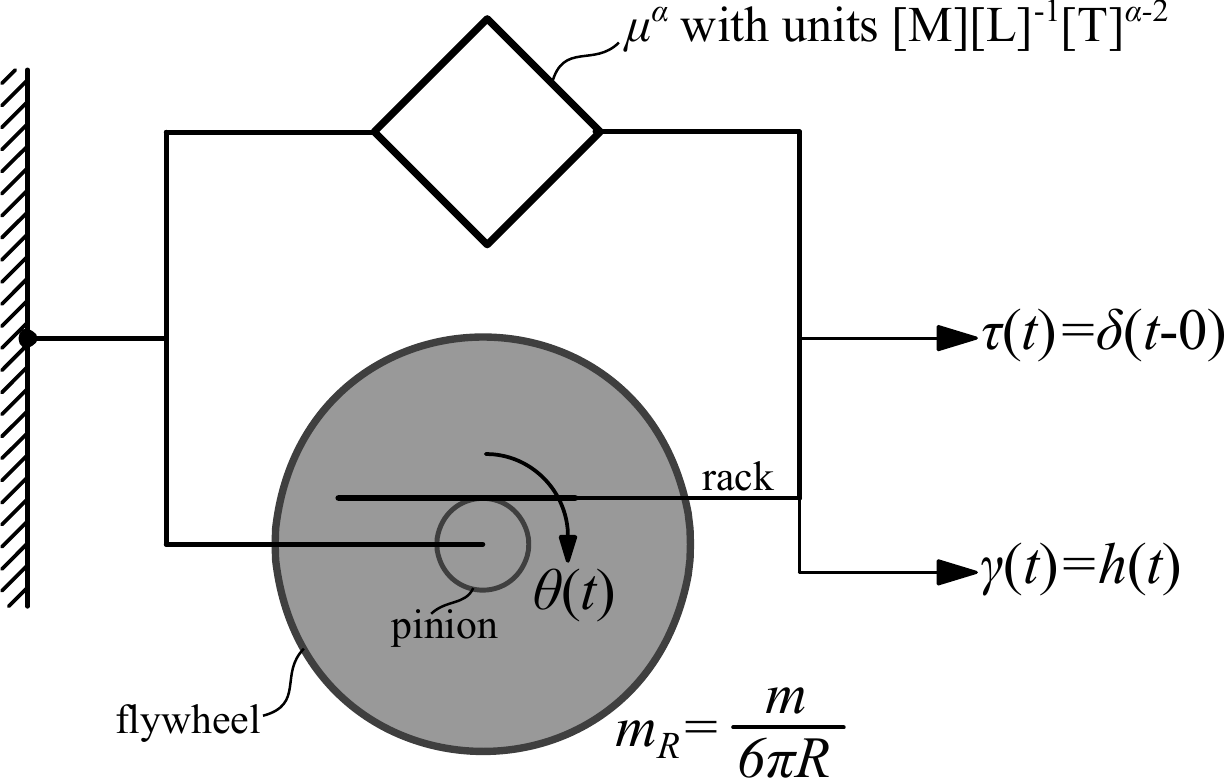}
\caption{A springpot--inerter parallel connection which is the mechanical analogue for Brownian motion of microspheres with mass $m$ and radius $R$ suspended in a Scott-Blair subdiffusive fluid with material constant $\mu_{\alpha}$ with units [M][L]$^{-1}$[T]$^{\alpha-2}$ (say Pa s$^{\alpha}$).}
\label{fig:Fig06}
\end{figure}

The rheological analogue for the Brownian motion of particles suspended in a Scott-Blair subdiffusive fluid is the springpot--inerter parallel connection shown in Fig. \ref{fig:Fig06} with constitutive law \cite{Makris2020}
\begin{equation}\label{eq:Eq46}
\tau(t)= \mu_{\alpha}\frac{\mathrm{d}^{\alpha}\gamma(t)}{\mathrm{d}t^{\alpha}} + m_R \frac{\mathrm{d}^2\gamma(t)}{\mathrm{d}t^2} , \quad \alpha\in\mathbb{R}^+
\end{equation}
The Laplace transform of Eq. (\ref{eq:Eq46}) is 
\begin{equation}\label{eq:Eq47}
\tau(s)=\mathcal{G}(s)\gamma(s)=(\mu_{\alpha}s^{\alpha}+m_R s^2)\gamma(s)
\end{equation}
where $\mathcal{G}(s)=\mu_{\alpha}s^{\alpha}+m_R s^2$ is the complex dynamic modulus of the springpot--inerter parallel connection, while the complex dynamic compliance is
\begin{equation}\label{eq:Eq48}
\mathcal{J}(s)=\frac{1}{\mathcal{G}(s)}=\frac{1}{\mu_{\alpha}s^{\alpha}+m_R s^2}=\frac{1}{m_R} \frac{1}{s^{\alpha}\left( s^{2-\alpha} +\frac{\mu_{\alpha}}{m_R} \right)}
\end{equation}
The inverse Laplace transform of Eq. (\ref{eq:Eq48}) is evaluated with the convolution integral \cite{Erdelyi1954}
\begin{figure}[t!]
\centering
\includegraphics[width=.7\linewidth, angle=0]{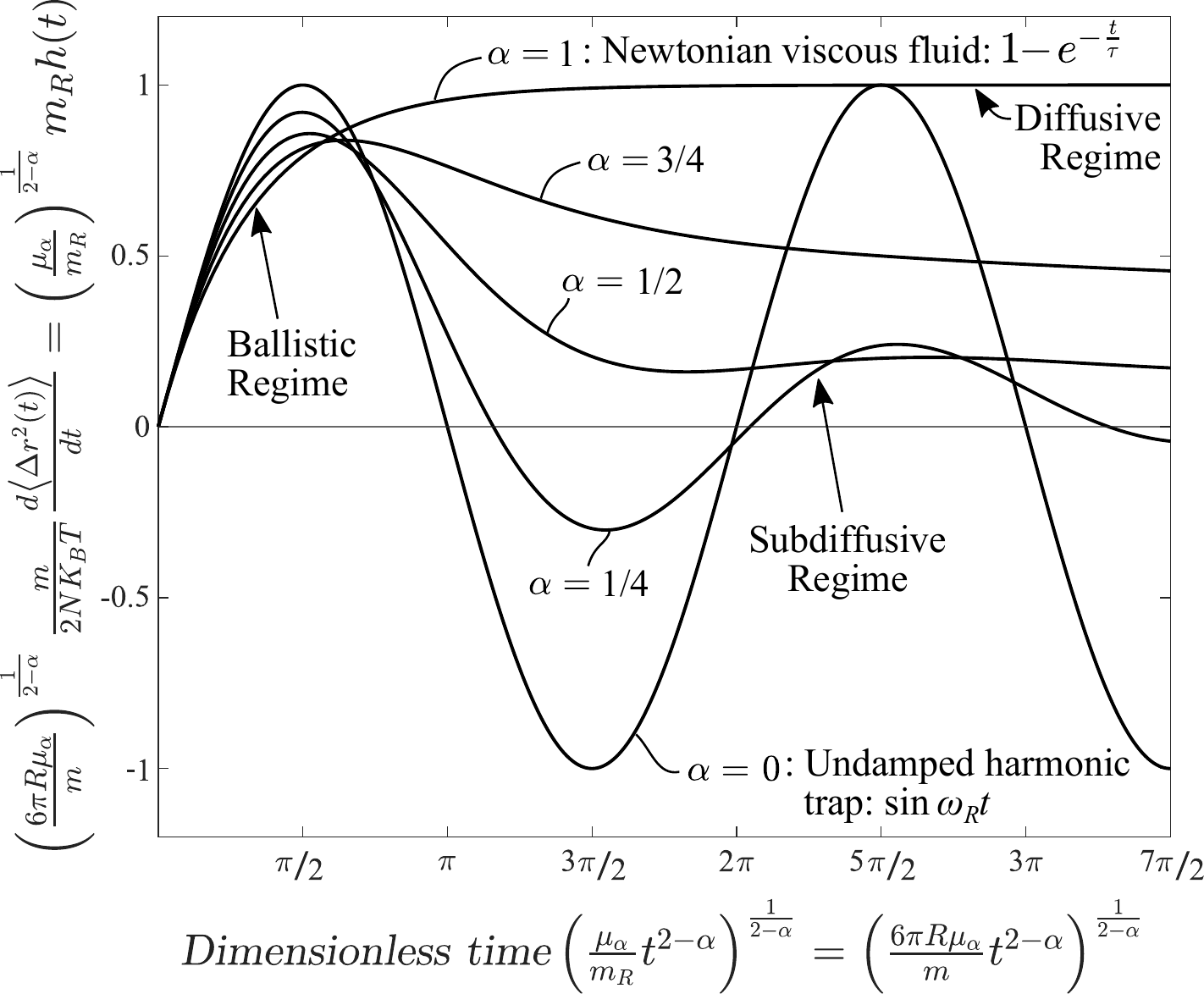}
\caption{Normalized time-derivative of the mean-square displacement of Brownian microspheres suspended in a fractional, subdiffusive Scott--Blair fluid with material constant $\mu_{\alpha}$ with units [M][L]$^{-1}$[T]$^{\alpha-2}$ for various values of the fractional exponent 0 $\leq \alpha \leq$ 1 as a function of the dimensionless time $\left( \frac{\mu_{\alpha}}{m_R}t^{2-\alpha}\right)^{\frac{1}{2-\alpha}}$, where $m_R=\frac{m}{6 \pi R}$ and $\tau = \frac{m}{6 \pi R \eta}$.}
\label{fig:Fig07}
\end{figure}
\begin{equation}\label{eq:Eq49}
h(t)=\mathcal{L}^{-1}\left\lbrace \mathcal{J}(s) \right\rbrace = \int_{0}^t f(t-\xi)g(\xi) \, \mathrm{d}\xi
\end{equation}
where 
\begin{equation}\label{eq:Eq50}
f(t)=\mathcal{L}^{-1}\left\lbrace \frac{1}{m_R} \frac{1}{s^{\alpha}} \right\rbrace = \frac{1}{m_R} \frac{1}{\Gamma(\alpha)} \frac{1}{t^{1-\alpha}} , \quad \alpha\in\mathbb{R}^+
\end{equation}
and
\begin{equation}\label{eq:Eq51}
g(t)  =\mathcal{L}^{-1}\left\lbrace \frac{1}{ s^{2-\alpha} +\frac{\mu_{\alpha}}{m_R} } \right\rbrace  = t^{1-\alpha} E_{2-\alpha,  \, 2-\alpha} \left( -\frac{\mu_{\alpha}}{m_R}t^{2-\alpha} \right)
\end{equation}
where $E_{2-\alpha,  \, 2-\alpha} \left( -\frac{\mu_{\alpha}}{m_R}t^{2-\alpha} \right)$ is the two-parameter Mittag--Leffler function defined by Eq. (\ref{eq:Eq45}). The function $g(t)$ expressed by Eq. (\ref{eq:Eq51}) is also known in rheology as the Rabotnov function, {\large $\varepsilon$}$_{1-\alpha}(-\lambda,  \, t)=t^{1-\alpha}E_{2-\alpha,  \, 2-\alpha}(-\lambda t^{2-\alpha})$ \cite{Rabotnov1980, Mainardi2010}. Upon substitution of the results of Eqs. (\ref{eq:Eq50}) and (\ref{eq:Eq51}) into the convolution integral given by Eq. (\ref{eq:Eq49}), the impulse response function of the springpot--inerter parallel connection shown in Fig. \ref{fig:Fig06} is merely the fractional integral of order $\alpha$ of the Rabotnov function given by Eq. (\ref{eq:Eq51}).
\begin{eqnarray}\label{eq:Eq52}
h(t) & =\frac{1}{m_R} \frac{1}{\Gamma(\alpha)} \int_0^t\frac{1}{(t-\xi)^{1-\alpha}}\xi^{1-\alpha}E_{2-\alpha,  \, 2-\alpha}\left( -\frac{\mu_{\alpha}}{m_R}\xi^{2-\alpha} \right) \, \mathrm{d}\xi \\ \nonumber 
& =\frac{1}{m_R} I^{\alpha} \left[ \varepsilon_{1-\alpha}\left( -\frac{\mu_{\alpha}}{m_R} , t \right) \right] = \frac{1}{m_R}tE_{2-\alpha,  \, 2}\left( -\frac{\mu_{\alpha}}{m_R}t^{2-\alpha} \right)
\end{eqnarray}
Substitution of the result of Eq. (\ref{eq:Eq52}) into Eq. (\ref{eq:Eq16}) together with $m_R=\frac{m}{6\pi R}$, yields the time derivative of the mean-square displacement of Brownian particles suspended in a Scott-Blair subdiffusive fluid
\begin{equation}\label{eq:Eq53}
\frac{\mathrm{d}\left\langle \Delta r^2 (t) \right\rangle}{\mathrm{d}t}  = \frac{N K_B T}{3 \pi R} h(t)  = \frac{2 N K_B T}{m} t E_{2-\alpha,  \, 2} \left( -\frac{6\pi R \mu_{\alpha}}{m} t^{2-\alpha}\right)
\end{equation}
The result of Eq. (\ref{eq:Eq53}) that was computed herein after calculating the impulse response function of the springpot--inerter parallel connection shown in Fig. \ref{fig:Fig06} in association with Eq. (\ref{eq:Eq16}) is identical to the first time derivative of Eq. (\ref{eq:Eq44}) derived by \cite{KobelevRomanov2000} and \cite{Lutz2001} after computing ensemble averages of the random Brownian process.

For the limiting case where $\alpha=$ 1, the Scott-Blair fluid becomes a Newtonian viscous fluid with $\mu_{\alpha}=\mu_1=\eta$ and Eq. (\ref{eq:Eq53}) reduces to 
\begin{equation}\label{eq:Eq54}
\frac{\mathrm{d}\left\langle \Delta r^2 (t) \right\rangle}{\mathrm{d}t} = \frac{2 N K_B T}{m} t E_{1 , \, 2}\left( -\frac{t}{\tau} \right)
\end{equation}
where $\tau = \frac{m}{6\pi R \eta}$ is the dissipation time. By virtue of the identity $E_{1 , \, 2}\left( -\frac{t}{\tau} \right)=\frac{\tau}{t}\left( 1-e^{-\nicefrac{t}{\tau}} \right)$, Eq. (\ref{eq:Eq54}) returns Eq. (\ref{eq:Eq10}) which was obtained by taking the time derivative of the mean-square displacement for Brownian motion in a memoryless Newtonian fluid derived by \cite{UhlenbeckOrnstein1930}.

At the other limiting case where $\alpha=$ 0, the Scott-Blair element becomes a Hookean elastic solid with $\mu_{\alpha}=\mu_{0}=G$ and Eq. (\ref{eq:Eq53}) gives
\begin{equation}\label{eq:Eq55}
\frac{\mathrm{d}\left\langle \Delta r^2 (t) \right\rangle}{\mathrm{d}t} = \frac{2 N K_B T}{m} t E_{2, \,2}\left( -\frac{6 \pi R G}{m} t^2 \right)
\end{equation}
where $\frac{6 \pi R G}{m}=\frac{G}{m_R}=\omega_0^2$. By virtue of the identity
\begin{equation}\label{eq:Eq56}
E_{2, \,2} \left( -\omega_0^2 t^2 \right) = \frac{\sinh \left( \mathrm{i} \omega_0 t \right)}{\mathrm{i} \omega_0 t } = \frac{1}{\omega_0 t}\sin\left( \omega_0 t \right) ,
\end{equation}
Eq. (\ref{eq:Eq55}) gives 
\begin{equation}\label{eq:Eq57}
\frac{\mathrm{d}\left\langle \Delta r^2 (t) \right\rangle}{\mathrm{d}t} = \frac{2 N K_B T}{m \omega_0} \sin (\omega_0 t)
\end{equation}
which is the special result of Eq. (\ref{eq:Eq26}) for Brownian motion in an undamped harmonic trap. Figure \ref{fig:Fig07} plots the normalized time-derivative of the mean-square displacement for Brownian motion in a subdiffusive Scott--Blair fluid
\begin{equation}\label{eq:Eq58}
\left( \frac{6 \pi R \mu_{\alpha}}{m} \right)^{\frac{1}{2-\alpha}}  \frac{m}{2 N K_B T} \frac{\mathrm{d}\left\langle \Delta r^2 (t)\right\rangle}{\mathrm{d}t}  =\left( \frac{\mu_{\alpha}}{m_R} \right)^{\frac{1}{2-\alpha}} m_R h(t)
\end{equation}
as a function of the dimensionless time $\left( \frac{\mu_{\alpha}}{m_R} t^{2-\alpha}\right)^{\frac{1}{2-\alpha}}$ with $m_R=\frac{m}{6\pi R}$ for various values of the fractional exponent $\alpha\in\mathbb{R}^+$.

\section{Summary}\label{sec:Sec07}

\begin{figure*}[t!]
\centering
\includegraphics[width=.75\linewidth, angle=0]{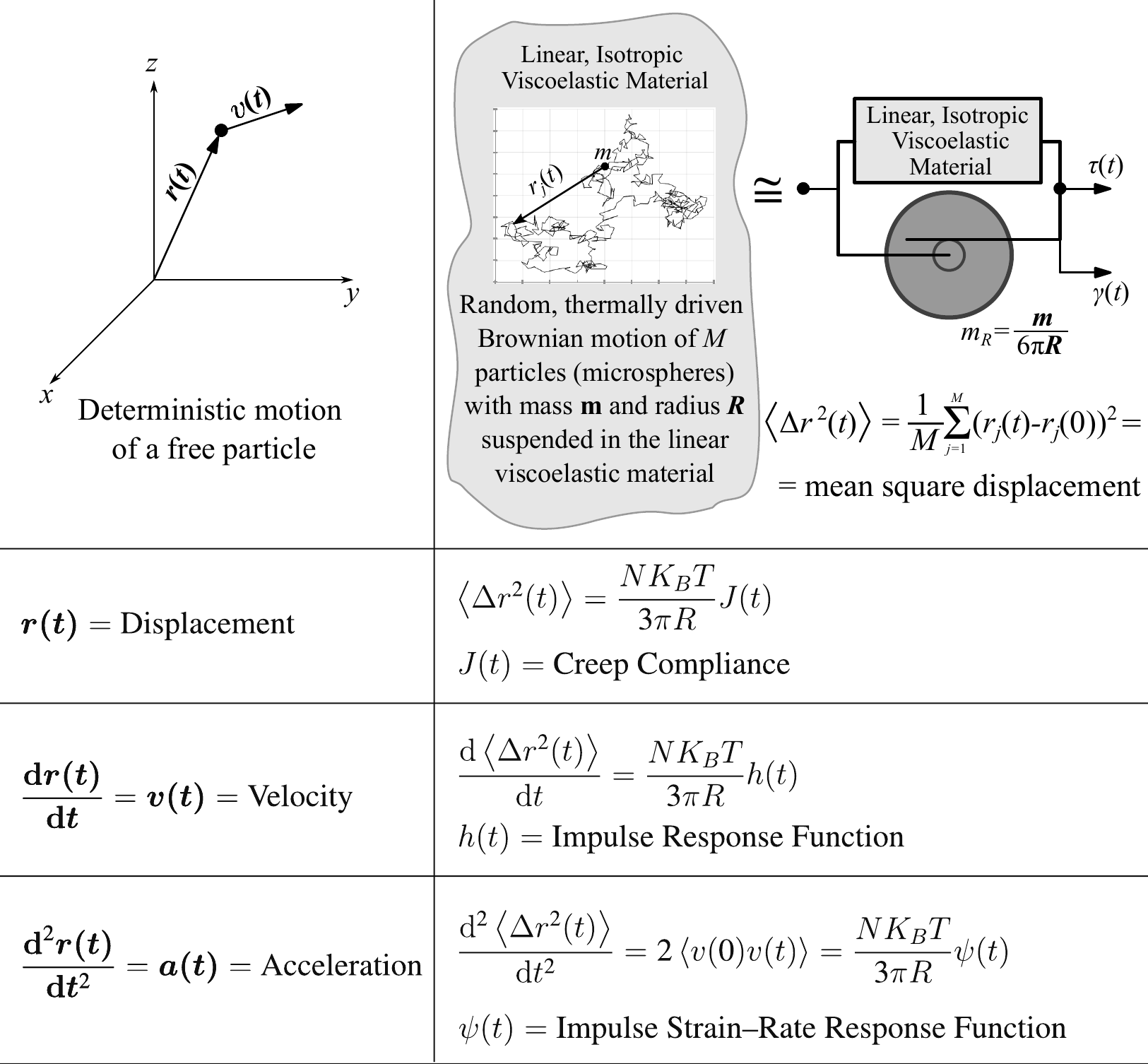}
\caption{In a analogous way that in Newtonian mechanics the deterministic motion of a free particle is described by its displacement $r(t)$, velocity $v(t)=\frac{\mathrm{d}r(t)}{\mathrm{d}t}$, and acceleration $a(t)=\frac{\mathrm{d}^2 r(t) }{\mathrm{d}t^2}$, the random, thermally driven Brownian motion of a collection of microspheres with mass $m$ and radius $R$ immersed in a linear, isotropic viscoelastic material is described by the deterministic creep compliance, $J(t)=\frac{3\pi R}{N K_B T}\left\langle \Delta r^{2} (t) \right\rangle$, impulse response function, $h(t)=\frac{3\pi R}{N K_B T}\frac{\mathrm{d}\left\langle \Delta r^{2} (t) \right\rangle}{\mathrm{d}t}$, and impulse strain-rate response function, $\psi(t)=\frac{3\pi R}{N K_B T}\frac{\mathrm{d}^2\left\langle \Delta r^{2} (t) \right\rangle}{\mathrm{d}t^2}$, of a mechanical network that is a parallel connection of the viscoelastic material (within which the microspheres are immersed) with an inerter with distributed inertance $m_R=\frac{m}{6\pi R}$. }
\label{fig:Fig08}
\end{figure*}

This paper builds upon past theoretical and experimental studies on Brownian motion and microrheology in association with a recently published viscous--viscoelastic correspondence principle for Brownian motion \cite{Makris2020} and shows that for all time-scales the time-derivative of the mean-square displacement, $\frac{\mathrm{d}\left\langle \Delta r^2 (t) \right\rangle}{\mathrm{d}t}$,
of Brownian microspheres with mass $m$ and radius $R$ suspended in any linear, isotropic viscoelastic material is identical to $\frac{N K_B T}{3 \pi R}h(t)$, where $h(t)$ is the impulse response function $($strain history $\gamma(t)$ due to an impulse stress $\tau(t)=\delta(t-0))$ of a linear mechanical network that is a parallel connection of the linear viscoelastic material (within which the Brownian microspheres are immersed) with an inerter with distributed inertance $m_R=\frac{m}{6 \pi R}$. The impulse response function $h(t)=\frac{3\pi R}{N K_B T}\frac{\mathrm{d}\left\langle \Delta r^2 (t) \right\rangle}{\mathrm{d}t}$ of the viscoelastic material--inerter parallel connection derived in this paper at the stress--strain level of the rheological analogue is essentially the response function $\chi(t)=\frac{h(t)}{6\pi R}$ of the Brownian particles expressed at the force--displacement level by Nishi \textit{et al.}\cite{NishiKilfoilSchmidtMacKintosh2018} after making use of the fluctuation--dissipation theorem. By employing the viscoelastic material--inerter rheological analogue we derive the mean-square displacement and its time-derivatives for Brownian particles suspended in a viscoelastic material described with a Maxwell element connected in parallel with a dashpot which captures the high-frequency viscous behavior and we show that for Brownian motion of micropaticles immersed in this class of fluid-like materials the impulse response function, $h(t)$ maintains a finite constant value in the long term.

With the introduction of the impulse response function $h(t)$ for Brownian motion, the paper uncovers that there is a direct analogy between the description of the deterministic motion of a free particle and the random motion of a collection of Brownian particles immersed in a linear viscoelastic material as illustrated in Fig. \ref{fig:Fig08}. This analogy shows that the random process of Brownian motion of a collection of particles can be fully described with the deterministic time--response functions of the viscoelastic material--inerter parallel connection.

\section*{References}
\bibliography{References} 
\bibliographystyle{iopart-num} 

\clearpage

\end{document}